\documentclass{aa}
\usepackage{graphicx}
\usepackage[varg]{txfonts}
\usepackage{hyperref}
\usepackage{booktabs}
\usepackage{xcolor}
\usepackage{soul}

\begin{document}
   \title{Estimating the heterogeneity of pressure profiles within a complete sample of 55 galaxy clusters: a Bayesian Hierarchical Model}
\authorrunning{Castagna et al.}
\titlerunning{Estimating the heterogeneity of pressure profiles within a complete sample of 55 galaxy clusters}
   \author{Fabio Castagna
          \inst{1,2}
          \and 
          Stefano Andreon\inst{1}
          \and 
          Marco Landoni\inst{3}
          \and 
          Alberto Trombetta\inst{2}
          }

    \institute{INAF-Osservatorio Astronomico di Brera, via Brera 28, 20121 Milano, Italy \\ \email{fabio.castagna@inaf.it} 
    \and
    Insubria University, Department of Pure and Applied Sciences, Computer Science Division, via Ottorino Rossi, 21100 Varese, Italy
    \and
    INAF-Osservatorio Astronomico di Brera, via Emilio Bianchi 46, 23807 Merate, Italy}

   \date{Received 18 April 2025
   / Accepted 7 August 2025
   }

    \abstract{
Galaxy clusters exhibit heterogeneity in their pressure profiles, even after rescaling, highlighting the need for adequately sized samples to accurately capture variations across the cluster population. We present a Bayesian hierarchical model that simultaneously fits individual cluster parameters and the underlying population distribution, providing estimates of the population-averaged pressure profile and the intrinsic scatter, as well as accurate pressure estimates for individual objects.
We introduce a highly flexible, low-covariance, and interpretable parameterization of the pressure profile based on restricted cubic splines. We model the scatter properly accounting for outliers, and we incorporate corrections for beam and transfer function, as required for Sunyaev-Zel'dovich (SZ) data. 
Our model is applied to the largest non-stacked sample of individual cluster radial profiles, extracted from SPT+\textit{Planck} Compton-$y$ maps. This is a complete sample of 55 clusters, with $0.05<z<0.30$ and $M_{500}>4\times 10^{14}M_\odot$, enabling subdivision into sizable morphological classes based on eROSITA data. Fitting is computationally feasible within a few days on a modern (2024) personal computer.
The shape of the population-averaged pressure profile, at our 250 kpc FWHM resolution, closely resembles the universal pressure profile, despite the flexibility of our model to accommodate alternative shapes, with a $\sim$12\% lower normalization, similar to what is needed to alleviate the tension between cosmological parameters derived from the cosmic microwave background and \textit{Planck} SZ cluster counts. Beyond $r_{500}$, our pressure profile is steeper than previous determinations. 
The intrinsic scatter is consistent with or lower than previous estimates, despite the broader diversity expected from our SZ selection.
Our flexible pressure modelization identifies a few clusters with non-standard concavity in their radial profiles but no outliers in amplitude. When dividing the sample by morphology, we find remarkably similar pressure profiles across classes, though regular clusters show evidence of lower scatter and a more centrally peaked profile compared to disturbed ones.
}


   \maketitle

\section{Introduction}

Galaxy clusters are the largest gravitationally bound objects in the universe and, as such, play a crucial role in studying cosmic evolution on the largest scales~\citep{Kravtsov2012}.
The Sunyaev-Zeldovich effect~\citep[SZ,][]{Sunyaev1970, Sunyaev1972} in the cosmic microwave background (CMB) is an ideal direct tracer of the gas pressure in the intra-cluster medium (ICM) of galaxy clusters. The hot gas trapped within the cluster's gravitational potential leaves an imprint on the microwave sky as its Compton free electrons scatter the photons of the CMB radiation.
The SZ effect depends directly on the integrated pressure along the line of sight, making it an ideal probe of the distribution of pressure across the ICM of galaxy clusters. This helps constrain models of structure formation for different cosmological scenarios~\citep{Pratt2019, Bocquet2024}.

Estimating the average pressure profile of a galaxy cluster population, along with its dispersion, provides valuable insights that extend beyond individual cluster studies. It enables the investigation of fundamental aspects of cosmology, astrophysics, and the large-scale structure of the universe, contributing to a more comprehensive understanding of cosmic evolution.
The scatter in the pressure profile within the population, together with other thermodynamic properties, is essential for analyzing the impact of non-gravitational physical processes such as gas cooling, supernova feedback~\citep{Tozzi2001, Kay2002}, and active galactic nuclei~\citep[AGN,][]{Gaspari2014}, as well as investigating hydrodynamic phenomena driven by shocks, turbulence, and bulk motions~\citep{Rasia2006, Vazza2009}.
The pressure profile can also be used to address more complex questions, such as those related to Degenerate Higher-Order Scalar-Tensor theories~\citep{Cardone2021}.
For all these reasons, many authors have studied the pressure profiles of multiple galaxy clusters~\citep{Arnaud2010, Planck2013, Sayers2013, Sayers2016, Bourdin2017, Romero2017, Ghirardini2019, Pointecouteau2021, Anbajagane2022, Melin2023, Oppizzi2023, Sayers2023, Hanser2023}, taking advantage of the increasing availability of data from SZ surveys, including \textit{Planck}, Bolocam, MUSTANG, SPT, ACT, and NIKA2.

When analyzing multiple galaxy clusters, it is important to recognize that they do not all share the same characteristics. Most studies in the literature scale profiles by mass and do not explicitly account for potential intrinsic variations among clusters at fixed mass, often treating such differences as part of the statistical scatter rather than modeling them directly.
However, it is crucial to examine the dispersion within the population and understand how individual clusters differ beyond the expected variations due to mass. 
Accounting for this heterogeneity is methodologically challenging and computationally demanding as it requires estimating the target quantity in a single-stage procedure for all clusters simultaneously. To properly incorporate the intrinsic scatter among distinct galaxy clusters, one must consider that the uncertainty in the average pressure profile arises not only from errors in individual clusters but also from their intrinsic differences.  
Normalizing different profiles according to their mass does not make them identical because the overall dispersion is larger than the spread caused by errors in the individual profiles.

To develop a model that inherently accounts for the statistical dispersion among distinct clusters, it is essential to recognize that the analysis operates at two interconnected levels: the top level, which describes the overall population and the distribution of individual clusters within it, and the bottom level, which refers to cluster-specific estimates, namely the properties of each individual galaxy cluster. For example, consider an analysis of human characteristics: the bottom level involves individual measurements such as mass, height, and income, while the top level represents the statistical distribution of these quantities across the entire population. From a statistical perspective, this structure can be naturally modeled using a Bayesian Hierarchical Model~\citep[BHM,][]{gelman1995bayesian, Andreon2015}.
The major obstacle in such an analysis is the computational cost, as the structure of the model and, consequently, the likelihood function, becomes increasingly complex as more clusters are added. This complexity scales more than linearly, making joint modeling significantly more computationally demanding than performing individual analyses and subsequently combining the results.

In this paper, we present a novel analysis of the pressure profile for a complete sample of 55 galaxy clusters.
The simultaneous fit is based on a BHM following a forward-modeling approach and relies on a Markov Chain Monte Carlo (MCMC) algorithm.

The paper is organized as follows: in Sect.~\ref{sec:methods}, we provide a detailed description of the methodology behind the analysis, including the data used, the pressure profile modelization, and the implementation of the BHM; in Sect.~\ref{sec:results}, we present the results of our analysis on a sample of 55 galaxy clusters and compare them with previously published studies; we discuss our findings in Sect.~\ref{sec:discussion} and conclude with the final remarks in Sect.~\ref{sec:conclusion}.

Throughout our work, we adopted a flat $\Lambda$CDM cosmology with $H_0=70 \mathrm{~km~s^{-1}~Mpc^{-1}}$, $\Omega_M=0.3$, and $\Omega_\Lambda=0.7$ to convert between observed and physical quantities. Throughout the paper, we adopt as summary measures for the posterior distribution of a parameter its median and the 68\% uncertainty defined by the corresponding percentiles of the distribution. 

\section{Methods} \label{sec:methods}

\subsection{Data}

\begin{figure}
\begin{center}
\begin{tabular}{c}
\includegraphics[width=.955\linewidth]{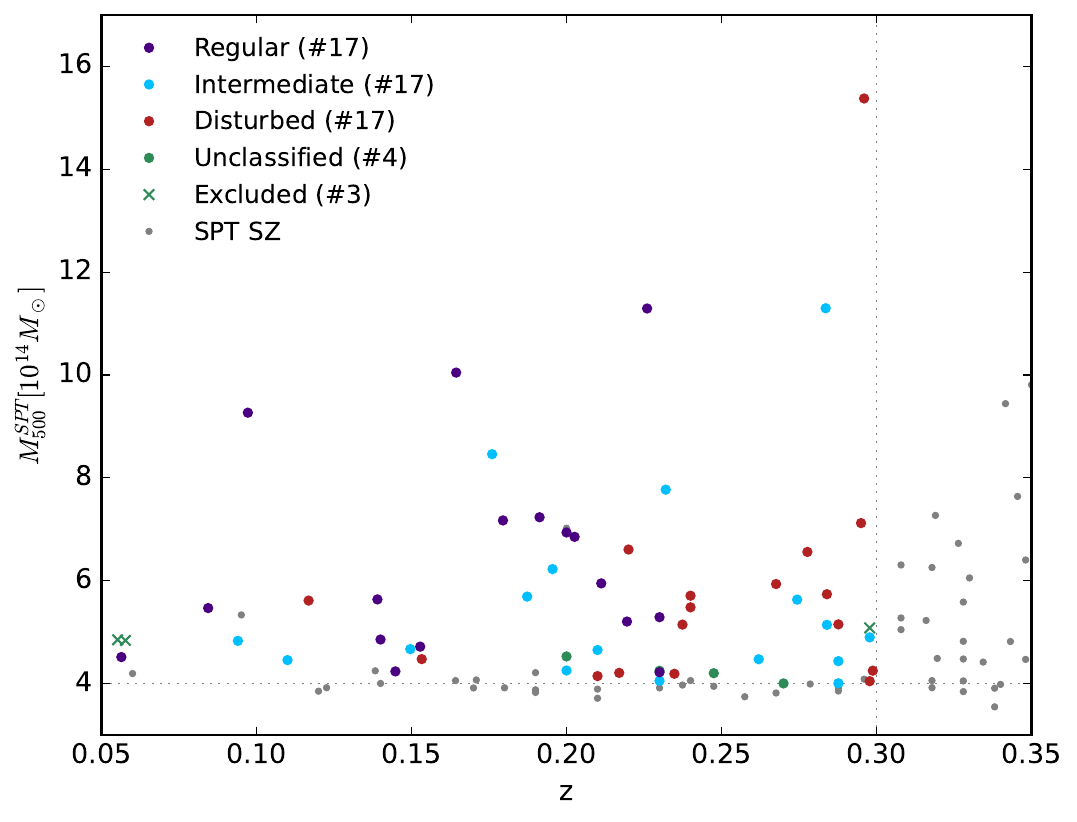}
\end{tabular}
\caption{Studied sample (colored points) and entire SPT catalog~\citep[gray points,][]{Bocquet2019}. Gray points within the redshift-mass selection area did not meet the S/N or footprint constraints. Crosses indicate galaxy clusters that were subsequently removed from our sample.} 
\label{fig:catalog}
\end{center}
\end{figure}

\begin{figure*}
\begin{center}
\begin{tabular}{cc}
\includegraphics[width=.45\linewidth]{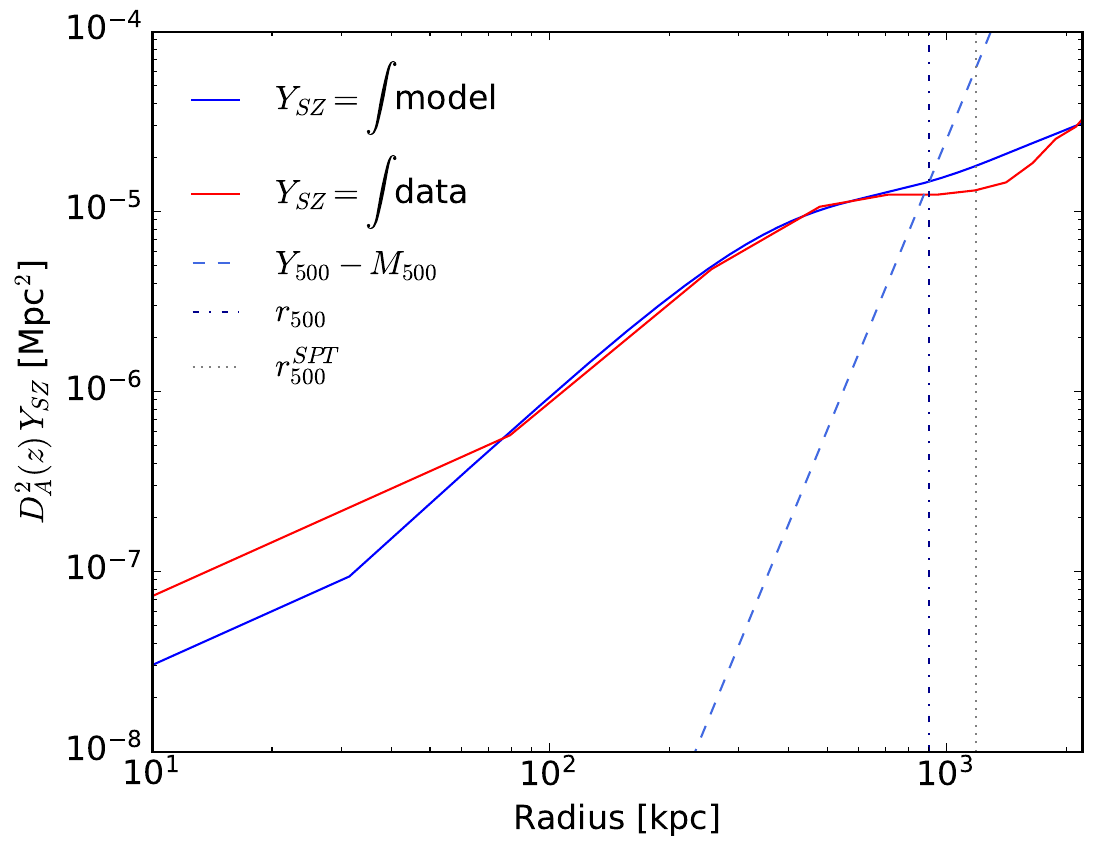} &
\includegraphics[width=.45\linewidth]{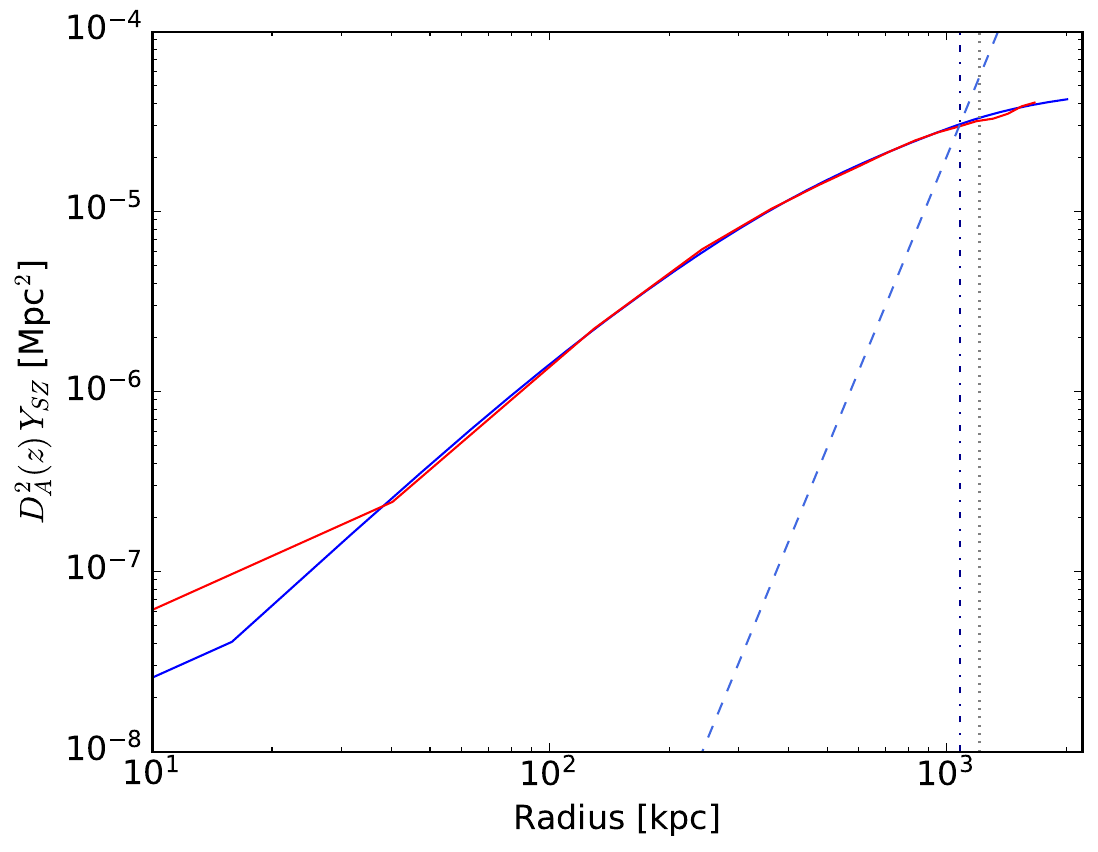}
\end{tabular}
\caption{Derivation of $r_{500}$ for two galaxy clusters, SPT-CLJ0051-4834 (left panel) and SPT-CLJ0328-5541 (right panel), shown as examples. 
Integrating the model ensures regularity in the case of noisy data (left panel) and does not affect high-quality data (right panel).}
\label{fig:r500_derivation}
\end{center}
\end{figure*}

From the South Pole Telescope (SPT) SZ galaxy cluster catalog~\citep{Bocquet2019}, we selected objects with $z < 0.3$, signal-to-noise ratio (S/N) $> 5$, $M_{500}^{SPT} > 4\times10^{14} M_\odot$, and that are not located at the boundary of the SPT map footprint~\citep[][see Fig.~\ref{fig:catalog}]{Bleem2022}. 
These criteria were guided by our goal of studying clusters with well-resolved, high-quality data. Out of the 58 selected clusters, in later sections we exclude two clusters (SPT-CLJ0431-6126 and SPT-CLJ2012-5649) because they are too extended in the sky, which makes excessive computation times for our hierarchical model and one cluster (SPT-CLJ0405-4916) because the background around it is oscillating instead of flattening (see Sect.~\ref{sec:exploratory}). Table~\ref{tab:individual} reports the sample of galaxy clusters analyzed.  

The raw Compton-$y$ map of each cluster was extracted from the Sanson-Flamsteed projection minimum-variance Compton-$y$ maps based on both SPT and \textit{Planck} data~\citep{Bleem2022}. Their radial profiles were computed in circular annuli with a width of 75 arcsec, which corresponds to the beam's full width at half maximum (FWHM), accounting for field boundaries and after flagging other sources and regions of lower quality. The chosen width ensures that the data covariance between radial bins is negligible. We adopted as centers the coordinates derived by \citet{Bocquet2019} and the point spread function (PSF) and transfer function provided with the Compton maps~\citep{Bleem2022}.
To estimate the errors of the profiles, we measured the scatter across profiles computed from random centers placed around each cluster. Radial profiles were extracted up to $r\sim17.5$ arcmin. At a round number close to the median redshift, $z=0.2$, the data are sampling the pressure profile with a 250 kpc scale FWHM, a scale much smaller than the cluster diameter at $\Delta=500$, typically 3 Mpc\footnote{The overdensity $r_{500}$ is defined as the radius within which the average density is $500$ times the critical density at the cluster's redshift.}.

We cross-matched our sample with the eROSITA catalog from the eRASS1 survey~\citep{Bulbul2024} to classify the clusters according to their morphological properties. We adopted the disturbance indicator $D_{\text{comb}}$ defined by \citet{Sanders2025}, which combines information on the shape and concentration of the cluster. Data were available for 51 out of the 55 galaxy clusters in our analysis. As shown in Fig.~\ref{fig:catalog}, we divided these clusters into three tertiles: $D_{\text{comb}}<0.02$ (dubbed regular), $0.02<D_{\text{comb}}<0.4$ (intermediate) and $D_{\text{comb}}>0.4$ (disturbed). 
Their morphological indicator is a combination of indices, not all of which account for the PSF, data noise, and redshift dependence, and which tend to be weighted toward the bright central regions of clusters, resulting in biases in the presence of a cool-core~\citep{Sanders2025}. 
Figure~\ref{fig:catalog} indicates that clusters classified as disturbed tend to have higher redshifts than those classified as relaxed.

\subsection{Pressure profile modelization} \label{sec:press_prof}

To model the pressure profile of galaxy clusters, we adopted restricted cubic splines~\citep{durrleman1989}.
These functions combine the flexibility of cubic interpolation between the innermost and outermost knots with linear regression at both extremities, thus avoiding undesired twists in the pressure profile at radii that are essentially unconstrained by data, namely well inside the beam FWHM and at radii where the S/N is vanishing.

The equation for a restricted cubic spline model with $K$ knots is:
\begin{equation}
    P(r) = \alpha + \beta_0 r+\sum_{j=1}^{K-2} \beta_j f(r, R_j),
\end{equation}
where
\begin{equation}
\begin{split}
    f(r,R_j) = \,& 
    \bigg[
    \big(r - R_j\big)^3_+ - \frac{R_K - R_j}{R_K - R_{K-1}} \big(r - R_{K-1}\big)^3_+ + \\
    & + \frac{R_{K-1} - R_j}{R_K - R_{K-1}} \big(r - R_K\big
    )^3_+ \bigg],
\end{split}
\end{equation}
where $(R_1, \dots, R_K)$ are the knots and the "+" notation indicates
\begin{equation}
u_+ = \begin{cases} u & \mbox{if} \ u \geq 0; \\ 0 & \mbox{otherwise}. \end{cases}
\end{equation}
Splines are computed in log-log space to avoid unphysical negative values for either radii or pressures.
To ensure a finite integral and to distinguish between a radially flat background and a radially flat cluster signal, we set an upper limit for the slope value beyond the largest knot, as done in previous works~\citep{Romero2018, Andreon2021}. In our analysis, we constrained this slope to be negative, which is less restrictive than in previous works.

Since clusters have different masses and sizes, we use the scaled pressure profiles $p(x)$ as defined in \citet{Arnaud2010} as follows:
\begin{equation}
    p(x)=\frac{P(r)}{P_{500}}\left[\frac{M_{500}}{3\times10^{14}M_\odot}\right]^{\alpha_P+\alpha'_P(x)},
    \label{eq:scaled_press}
\end{equation}
where $x=r/r_{500}$,  $\alpha_P\simeq0.12$, and
\begin{equation}
    \alpha'_P(x)=0.10-\left(\alpha_P+0.10\right)\frac{\left(x/0.5\right)^3}{1+\left(x/0.5\right)^3}.
\end{equation}
$P_{500}$, the characteristic pressure, is computed according to the following equation:
\begin{equation}
    P_{500}=\frac{3}{8\pi}\left[\frac{500\,G^{-1/4}H(z)^2}{2}\right]^{4/3}\frac{\mu}{\mu_e}\,f_B\,M_{500}^{2/3},
    \label{eq:P500}
\end{equation}
where $H(z)=H_0\sqrt{\Omega_M\left(1+z\right)^3+\Omega_\Lambda}$ is the Hubble parameter at a given redshift, $G$ is the gravitational constant, $f_B=\Omega_B/\Omega_M=0.175$ is the mean baryon fraction in the universe, $\mu=0.59$ is the mean molecular weight, and $\mu_e=1.14$ is the mean molecular weight per free electrons. 

\subsection{Likelihood function computation for the i-th cluster}

The three-dimensional pressure profile of Sect.~\ref{sec:press_prof} is numerically integrated along the line of sight through an Abel transform to obtain a two-dimensional map of the Compton-$y$ parameter, which is then convolved to account for the beam and transfer function. 
Finally, the radial profile is extracted from the filtered Compton-$y$ parameter map, converted to the surface brightness unit, and compared to the observed data within a Bayesian framework, yielding the pressure profile parameters and their associated errors.
We added a pedestal parameter to the surface brightness profile to account for a non-zero background level, adopting a Gaussian distribution centered on 0 with $\sigma=10^{-6}$ as prior. For all these operations, we used \texttt{PreProFit}, as detailed in \citet{Castagna2019}. The updated code, which allows users to adopt a restricted cubic spline model, as well as a generalized Navarro, Frenk \& White (gNFW) profile~\citep{Nagai2007}, a piecewise power-law, or a cubic spline model, is publicly available on GitHub\footnote{\url{https://github.com/fcastagna/preprofit}}.

\begin{figure}
\centering
\begin{center}
\begin{tabular}{c}
\includegraphics[width=.955\linewidth]{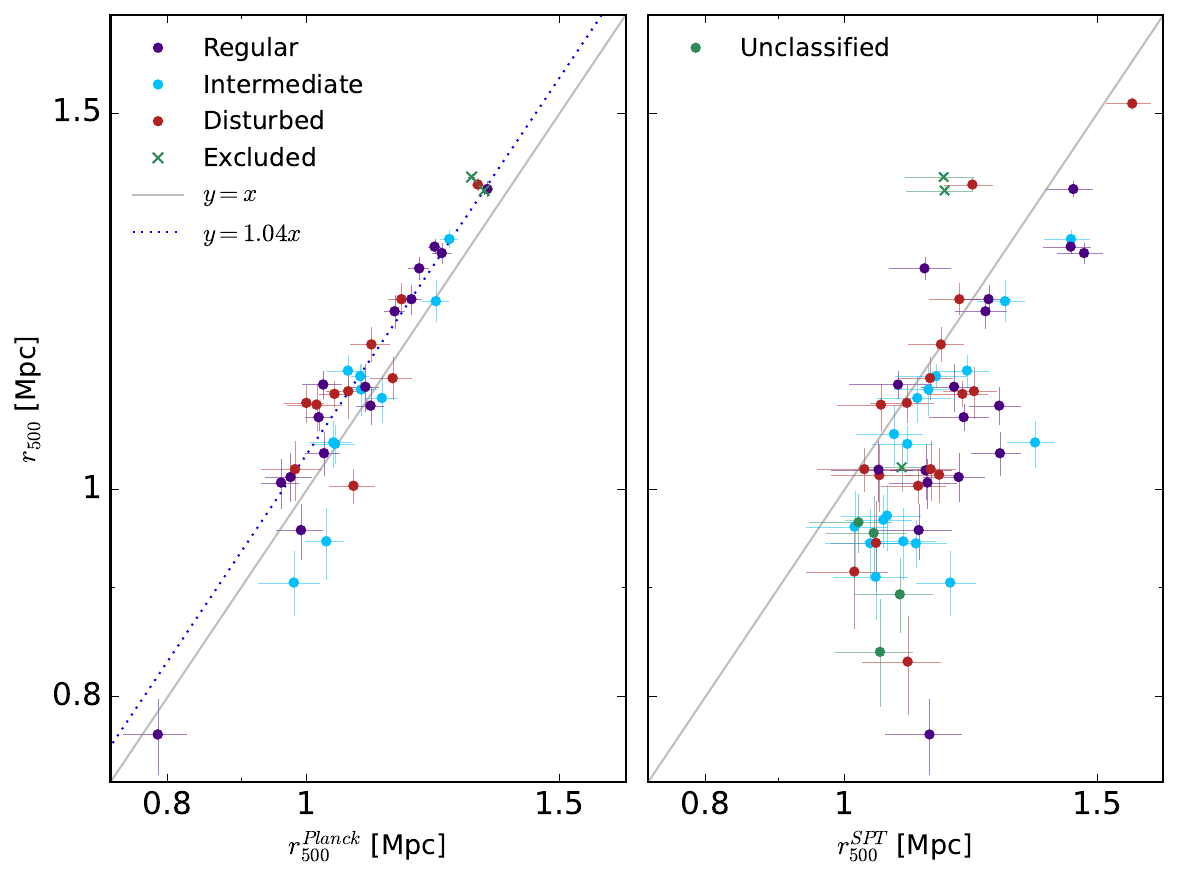}
\end{tabular}
\caption{Comparison of our estimates of $r_{500}$ (y-axis) with \textit{Planck} for the 38 clusters in common (left panel) and with SPT for the entire sample (right panel). Our $r_{500}$ values are strongly correlated with the \textit{Planck} ones, while a larger scatter is observed when compared with SPT. No morphology-dependent trend is observed.
} 
\label{fig:r500}
\end{center}
\end{figure} 

\subsection{Mass definition} \label{subsec:masses}

The $M_{500}^{SPT}$ values in the SPT catalog~\citep{Bocquet2019} are unreliable because the derived $r_{500}^{SPT}$ values are, in some cases, manifestly too large or too small compared to the angular size of the cluster, regardless of whether these values are corrected by the 0.8 factor proposed in \citet{Tarrio2019}, as illustrated below.
We therefore estimated the cluster's $M_{500}$ mass from its SZ flux, $Y_{500}$, using
\begin{equation}
    E^{-2/3}(z) \left[ \frac{D_A^2 Y_{500}}{10^{-4} \mathrm{Mpc}^2} \right] = 10^{-0.19} \left[ \frac{M_{500}}{6\times 10^{14} M_\odot} \right]^{1.79}, \label{eq:Y500-M500}
\end{equation}
where $E(z)=H(z)/H_0$ is the dimensionless Hubble parameter and $D_A$ is the angular diameter distance~\citep{Planck2014}, as illustrated in Fig.~\ref{fig:r500_derivation} for two clusters. We adopted $r_{500}$ values derived from the fitted pressure profile model rather than from the data because the model enforces regularity in the case of noisy data (e.g., see left panel) and does not harm high-quality data (e.g., see right panel). $Y_{500}$ and $M_{500}$ values for each cluster are reported in Table~\ref{tab:individual}.

Figure~\ref{fig:r500} compares our $r_{500}$ values with the $r_{500}^{\textit{Planck}}$ values from the \textit{Planck} catalog~\citep{Planck2016} in the left panel for the 38 objects in common, and with the $r_{500}^{SPT}$ values from the SPT catalog~\citep{Bocquet2019} in the right panel for the entire sample. The $r_{500}^{\textit{Planck}}$ values are derived from a measurement at $5r_{500}$, scaled down by an amount fixed by the \textit{Planck} team assuming the pressure profile of \citet{Arnaud2010}.
The $r_{500}^{SPT}$ values, on the other hand, are inferred in \citet{Bocquet2019} from the cluster detection probability. 
Our $r_{500}$ values are in good agreement with \textit{Planck} values, with most of them tightly aligned along the $r_{500}=1.04r_{500}^{\textit{Planck}}$ relation (blue dotted line).
On the other hand, we observe no agreement between our $r_{500}$ and $r_{500}^{SPT}$, which justifies the decision to recompute these values through $Y_{500}$.
In particular, the $r_{500}^{SPT}$ values show a large scatter because the spatial Compton-$y$ signal assumed an overly constrained profile model, as earlier mentioned and detailed in Sect.~\ref{sec:exploratory}.

\subsection{Hierarchical model for the population of clusters}

To model the variety of pressure profiles of clusters with the same mass, we used a two-level BHM.
This approach allows us to estimate the average pressure profile and the intrinsic scatter for a population of galaxy clusters. The top level models the population, while the bottom level models the individual clusters. 
The population-averaged pressure profile is modeled (non-parametrically) with a restricted cubic spline with $n_k$ knots. At each knot, the scatter of the individual profiles from the average is modeled with a Student's $t$ distribution with 10 degrees of freedom. We preferred a Student's $t$ over a Gaussian because it is less sensitive to outliers~\citep[e.g.,][]{Andreon2010}. However, we also performed an additional analysis using a Gaussian distribution for the scatter, which yielded very similar results.

In detail, at the $k$-th knot of $i$-th cluster, the individual pressures $lgP_{k,i}$ can scatter around the population-averaged pressure $lgP_k$ as
\begin{equation}
    lgP_{k,i}\sim t_{\nu=10}\left( \mu=lgP_k, \sigma=\sigma_{int,k}\sqrt{\frac{\nu-2}{\nu}}\right).
    \label{eq:lgpki}
\end{equation}
The coefficient in front of the scale parameter ensures that $Var\left(t_\nu\right)=\sigma_{int,k}^2$, allowing us to correctly interpret the $\sigma_{int,k}$ parameters as estimates of the intrinsic scatter. $lgP$ is a shorthand for $\log_{10}(p(x))$.

Our prior of the population-averaged pressure profile at the $k$-th knot is a Gaussian with a large $\sigma$ (0.5 dex) centered on the universal pressure profile~\citep[UPP,][]{Arnaud2010}: 
\begin{equation} 
    lgP_k \sim N\left(\mu=lgP^\mathrm{UPP}(x_k), \sigma=0.5\right).
\end{equation} 

The prior on the scatter is a uniform distribution over a wide range (from 0 to 1 dex), largely encompassing all plausible values:
\begin{equation}
    \sigma_{int,k} \sim UC\left(0, 1\right).
\end{equation}

Our model has $(2\times n_k)$ parameters describing the population, namely the average pressures at the $n_k$ knots and the dispersions around them, plus ($n_k$+1) parameters per cluster, namely the individual pressures at the $n_k$ knots and the pedestal parameter.

\subsection{Model and MCMC settings}

Throughout our analyses, we set the number of knots to $n_k=5$ and placed them at fixed locations: $\left[ 0.1, 0.4, 0.7, 1, 1.3 \right] \times r_{500}$. With this radial scaling, we assume that the scatter is a function of $r_{500}$, as in previous works~\citep[e.g.,][]{Ghirardini2019}, rather than, for example, radii that do not scale with mass. The spacing of the knots is chosen to be larger than the width of the radial binning to avoid strong covariances between radially adjacent estimates, while the innermost radius is set to prevent undersampling the PSF FWHM.
The MCMC estimations were performed using the PyMC Python package\footnote{\url{https://www.pymc.io/welcome.html}}~\citep{Patil2010} with the default sampling methods.
We ran 8000 iterations with 24 walkers, discarding the first half iterations as a burn-in period. We then compared the posterior distributions of all walkers and removed some walkers that had not fully converged.

\section{Results} \label{sec:results}

\begin{figure}
\begin{center}
\begin{tabular}{c}
\includegraphics[width=.89\linewidth]{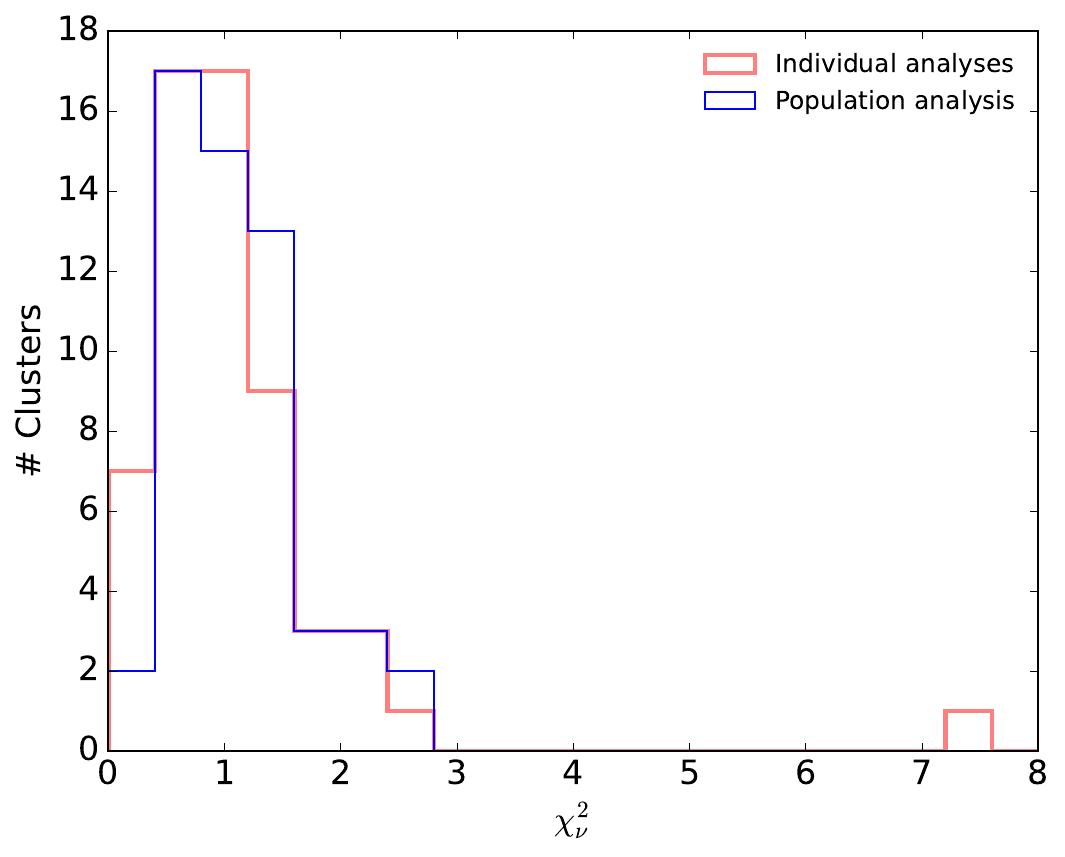}
\end{tabular}
\caption{Distribution of the Reduced Chi-Square $\chi^2_{\nu=6}$ of the 58 individual fits and the population fit of 55 clusters. The red outlier with $\chi^2_{\nu}>7$ is SPT-CLJ0405-4916, which is characterized by a highly oscillating surface brightness distribution in its outermost region, while our background model, the pedestal, is a constant. For this reason, we excluded it from the population analysis.
} 
\label{fig:chisq}
\end{center}
\end{figure}

\subsection{Exploratory analysis and cluster characterization} \label{sec:exploratory}

We start by fitting each individual galaxy cluster independently of the other ones using a standard Bayesian analysis with uniform priors for the pressure values to assess whether our chosen model of the pressure profile is adequate for the used data. 
Figure~\ref{fig:chisq} shows the distribution of the Reduced Chi-Square $\chi_{\nu=6}^2$ for this exploratory analysis (red histogram), compared with the analogous measures for the subsequent population analysis (blue histogram). 
This goodness-of-fit evaluation is performed by comparing the median surface brightness profile, obtained by computing the median value across all posterior profiles at each radial bin, with the observed data, and therefore does not provide the minimal $\chi_\nu^2$. 
All individual fits, except one, show acceptable $\chi_\nu^2$ values, suggesting that the adopted spline with five knots is sufficiently flexible to describe the pressure profiles of the studied clusters given the data quality. The object with a large $\chi_\nu^2$ is SPT-CLJ0405-4916, and its value is due to an oscillating background level due to contamination. For this reason, SPT-CLJ0405-4916 has been removed from the sample.
The results of the population fit show acceptable $\chi_\nu^2$ values, suggesting that the population analysis presented in the following section appropriately models all clusters in the sample.

\begin{figure}
\begin{center}
\begin{tabular}{c}
\includegraphics[width=.92\linewidth]{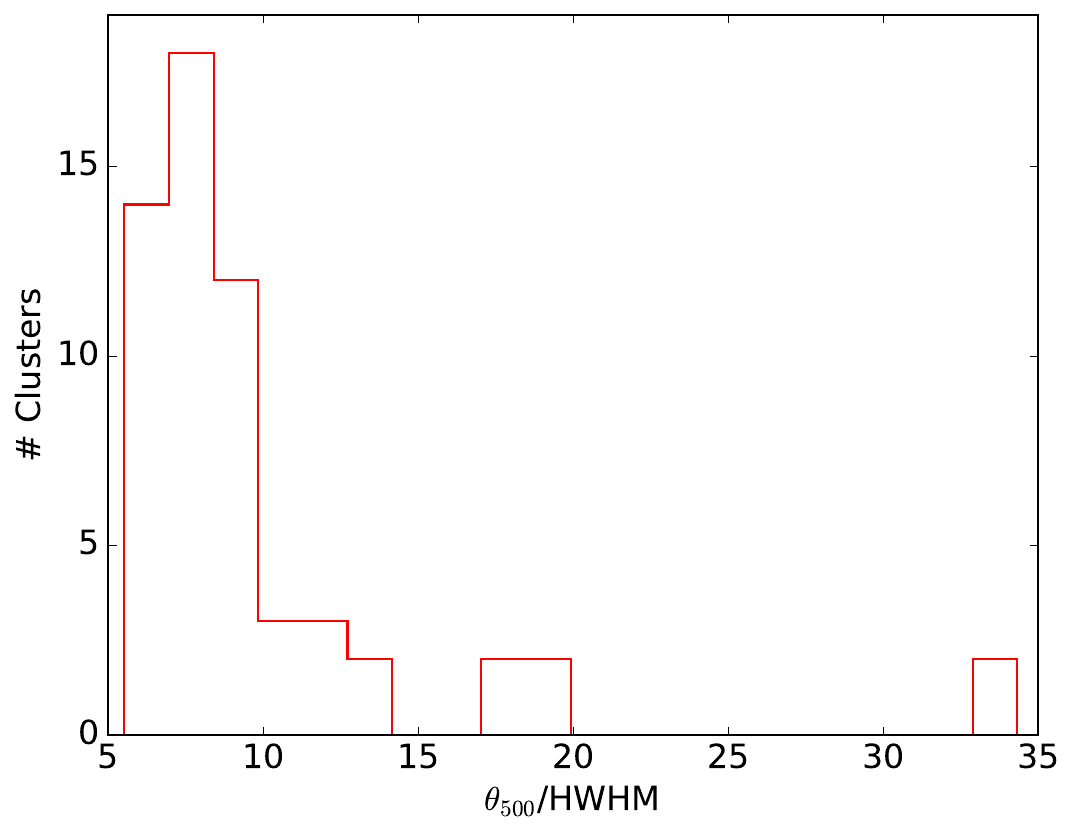}
\end{tabular}
\caption{Distribution of the number of resolution elements, $\theta_{500}/$HWHM, across the initial sample of 58 objects. Because of our careful sample selection in mass and redshift, all clusters have $\theta_{500}/$HWHM > 5, i.e., are well-resolved. The two clusters with $\theta_{500}/$HWHM $> 30$ are excluded from later analyses because of the high computational time required to include them in the hierarchical model.} 
\label{fig:theta_500}
\end{center}
\end{figure}

Figure~\ref{fig:theta_500} shows the distribution of the number of resolution elements for all 58 clusters, defined as the ratio between $\theta_{500}$, namely the characteristic radius $r_{500}$ in angular units, and the beam's half width at half maximum (HWHM), which in our case equals 0.625 arcmin. Because of our selection based on mass and redshift, all clusters have $\theta_{500}/$HWHM $> 5$, i.e., are well-resolved, making the assumption used in \citet{Bocquet2019} to derive their (SZ) mass invalid for our sample and justifying our redetermination of it.
Two objects (SPT-CLJ2012-5649 and SPT-CLJ0431-6126), with $\theta_{500}/$HWHM $> 30$, are so extended that we chose to exclude them from our sample for computational reasons: including them would considerably increase the computation time of the hierarchical model.

\begin{figure}
\begin{center}
\begin{tabular}{c}
\includegraphics[width=.955\linewidth]{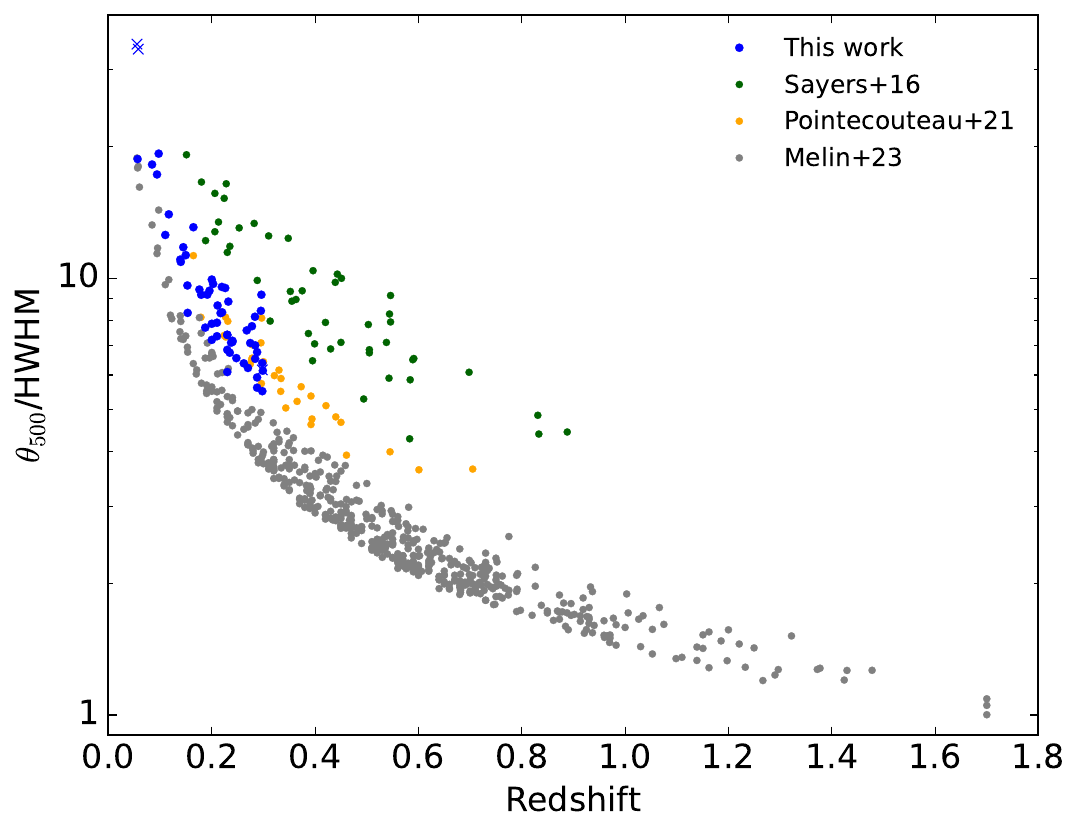}
\end{tabular}
\caption{Resolution of galaxy clusters in our sample and in works that analyze galaxy clusters using only SZ instruments~\citep{Sayers2016, Pointecouteau2021, Melin2023}. Compared to other studies, our selection focuses on well-resolved clusters.}
\label{fig:catalogs}
\end{center}
\end{figure}

Figure~\ref{fig:catalogs} shows how our selection resulted in a sample of more resolved clusters compared to other works in the literature that analyzed samples of clusters using SZ data~\citep{Sayers2016, Pointecouteau2021, Melin2023}.

\begin{figure}
\begin{center}
\begin{tabular}{c}
\includegraphics[width=.95\linewidth]{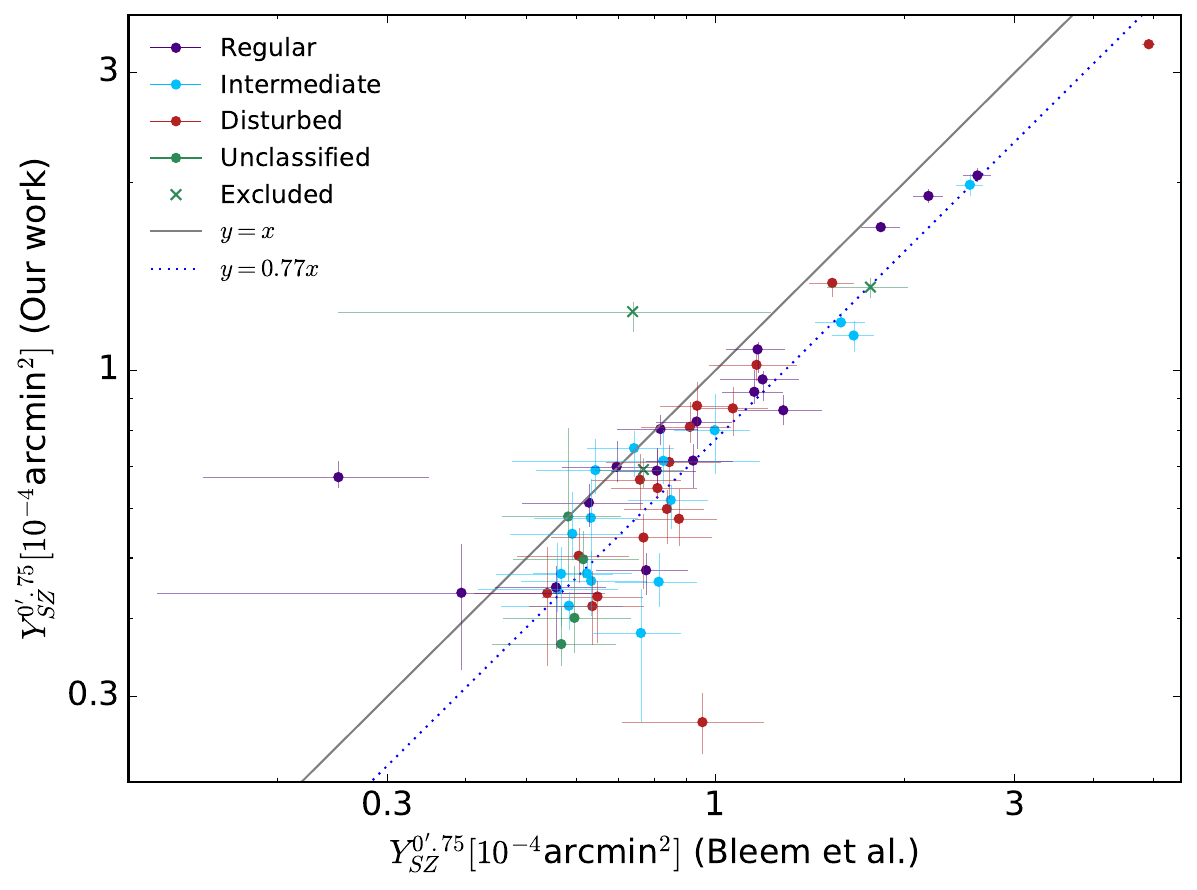}
\end{tabular}
\caption{Integrated Compton parameter $Y^{0^\prime.75}_{SZ}$ computed from our fit to individual clusters (median with 68\% uncertainty) vs values from \citet[][mean with 1-$\sigma$ error]{Bleem2015}. 
Despite the strongly constrained shape that they assumed for their estimation method, there is a clear trend between the two estimates, although with an offset.}
\label{fig:y75}
\end{center}
\end{figure} 

Figure~\ref{fig:y75} compares our integrated Compton parameters within 0.75 arcmin, $Y^{0'.75}_{SZ}$, with the values published by \citet{Bleem2015} for the sample of 58 clusters. The determination of \citet{Bleem2015} assumes that the cluster profile follows a $\beta$-model,
and uses shallower data. Apart from the fact that their values are larger by approximately 23\%, likely due to the overly constrained model adopted, there is general agreement between the two determinations. Our values are reported in Table~\ref{tab:individual}.

\subsection{Population analysis} \label{sec:population}

\begin{table*}[ht]
\centering\caption{Population-averaged pressure and intrinsic scatter values for the entire studied sample and for the morphological classes.}
\footnotesize{
\begin{tabular}{rrrrrrrrr}
\toprule
    & \multicolumn{2}{c}{All clusters} & \multicolumn{2}{c}{Regular} & \multicolumn{2}{c}{Intermediate} & \multicolumn{2}{c}{Disturbed} \\
$r/r_{500}$ & $lgP_k$ & $\sigma_{int,k}$ & $lgP_k$ & $\sigma_{int,k}$ & $lgP_k$ & $\sigma_{int,k}$ & $lgP_k$ & $\sigma_{int,k}$ \\
& & [dex] & & [dex] & & [dex] & & [dex] \\
\midrule 
0.1 & 0.88$^{+0.05}_{-0.05}$ & 0.20$^{+0.05}_{-0.04}$ & 0.94$^{+0.07}_{-0.08}$ & 0.26$^{+0.10}_{-0.07}$ & 0.86$^{+0.09}_{-0.09}$ & 0.14$^{+0.11}_{-0.08}$ & 0.71$^{+0.10}_{-0.10}$ & 0.10$^{+0.11}_{-0.06}$ \\ 0.4 & 0.15$^{+0.02}_{-0.02}$ & 0.09$^{+0.02}_{-0.02}$ & 0.14$^{+0.02}_{-0.03}$ & 0.05$^{+0.03}_{-0.02}$ & 0.15$^{+0.03}_{-0.04}$ & 0.09$^{+0.04}_{-0.03}$ & 0.17$^{+0.04}_{-0.05}$ & 0.13$^{+0.04}_{-0.03}$ \\ 0.7 & -0.37$^{+0.02}_{-0.02}$ & 0.08$^{+0.03}_{-0.03}$ & -0.40$^{+0.03}_{-0.03}$ & 0.04$^{+0.04}_{-0.02}$ & -0.41$^{+0.06}_{-0.06}$ & 0.15$^{+0.07}_{-0.06}$ & -0.31$^{+0.05}_{-0.05}$ & 0.09$^{+0.06}_{-0.04}$ \\ 1.0 & -0.75$^{+0.04}_{-0.04}$ & 0.14$^{+0.03}_{-0.03}$ & -0.75$^{+0.05}_{-0.05}$ & 0.12$^{+0.05}_{-0.04}$ & -0.75$^{+0.08}_{-0.08}$ & 0.12$^{+0.09}_{-0.07}$ & -0.72$^{+0.08}_{-0.08}$ & 0.20$^{+0.08}_{-0.06}$ \\ 1.3 & -1.29$^{+0.05}_{-0.05}$ & 0.19$^{+0.05}_{-0.04}$ & -1.23$^{+0.06}_{-0.07}$ & 0.16$^{+0.07}_{-0.05}$ & -1.36$^{+0.11}_{-0.14}$ & 0.27$^{+0.16}_{-0.10}$ & -1.39$^{+0.10}_{-0.14}$ & 0.31$^{+0.15}_{-0.09}$ \\
\bottomrule
\end{tabular}}
\label{tab:population}
\end{table*}

\begin{table*}[ht]
\centering
\caption{Sample of galaxy clusters and estimated parameters in the population analysis. The full table is available at CDS (add URL when available).}
\resizebox{\textwidth}{!}{\begin{tabular}{lrrrrrrrrrrr}
\toprule
Name & RA & Dec & $z$ & $M_{500}$ & $Y_{500}$ & $Y_{SZ}^{0'75}$& $lgP_{0,i}$ & $lgP_{1,i}$ & $lgP_{2,i}$ & $lgP_{3,i}$ & $lgP_{4,i}$ \\ 
& [deg] & [deg] & & [10$^{14}M_\odot$] & [10$^{-4}$ Mpc$^2$] & [10$^{-4}$ arcmin$^2$] & & & & & \\
\midrule
SPT-CLJ0022-4144 & 5.5489 & -41.7366 & 0.298 & 2.96$^{+0.51}_{-0.46}$ & 0.20$^{+0.05}_{-0.05}$ & 0.43$^{+0.03}_{-0.07}$ & 0.87$^{+0.18}_{-0.19}$ & 0.15$^{+0.07}_{-0.08}$ & -0.37$^{+0.07}_{-0.08}$ & -0.74$^{+0.12}_{-0.12}$ & -1.34$^{+0.14}_{-0.18}$ \\ SPT-CLJ0027-5015 & 6.8228 & -50.2524 & 0.145 & 4.60$^{+0.21}_{-0.21}$ & 0.42$^{+0.03}_{-0.03}$ & 0.67$^{+0.04}_{-0.03}$ & 0.79$^{+0.12}_{-0.15}$ & 0.11$^{+0.05}_{-0.06}$ & -0.33$^{+0.06}_{-0.05}$ & -0.70$^{+0.10}_{-0.10}$ & -1.25$^{+0.12}_{-0.15}$ \\ SPT-CLJ0051-4834 & 12.7905 & -48.5776 & 0.187 & 2.54$^{+0.26}_{-0.27}$ & 0.15$^{+0.02}_{-0.03}$ & 0.46$^{+0.05}_{-0.04}$ & 0.94$^{+0.19}_{-0.19}$ & 0.19$^{+0.07}_{-0.07}$ & -0.41$^{+0.07}_{-0.09}$ & -0.83$^{+0.11}_{-0.13}$ & -1.34$^{+0.15}_{-0.20}$ \\ SPT-CLJ0118-5638 & 19.5385 & -56.6339 & 0.210 & 3.68$^{+0.41}_{-0.37}$ & 0.29$^{+0.04}_{-0.04}$ & 0.44$^{+0.08}_{-0.10}$ & 0.79$^{+0.16}_{-0.20}$ & 0.10$^{+0.07}_{-0.08}$ & -0.38$^{+0.06}_{-0.07}$ & -0.70$^{+0.09}_{-0.08}$ & -0.99$^{+0.07}_{-0.08}$ \\ SPT-CLJ0124-5937 & 21.1988 & -59.6255 & 0.210 & 2.97$^{+0.35}_{-0.34}$ & 0.20$^{+0.03}_{-0.03}$ & 0.46$^{+0.05}_{-0.06}$ & 0.88$^{+0.17}_{-0.18}$ & 0.11$^{+0.06}_{-0.07}$ & -0.39$^{+0.06}_{-0.08}$ & -0.78$^{+0.12}_{-0.12}$ & -1.26$^{+0.15}_{-0.18}$ \\ SPT-CLJ0143-4452 & 25.8853 & -44.8741 & 0.270 & 3.37$^{+0.32}_{-0.32}$ & 0.25$^{+0.04}_{-0.04}$ & 0.36$^{+0.06}_{-0.03}$ & 0.79$^{+0.16}_{-0.19}$ & 0.12$^{+0.06}_{-0.07}$ & -0.34$^{+0.09}_{-0.07}$ & -0.72$^{+0.13}_{-0.12}$ & -1.22$^{+0.14}_{-0.16}$ \\ \bottomrule
\end{tabular}
\label{tab:individual}
}
\end{table*}

\begin{figure}[t]
\centering
\includegraphics[width=.48\textwidth, page=1]{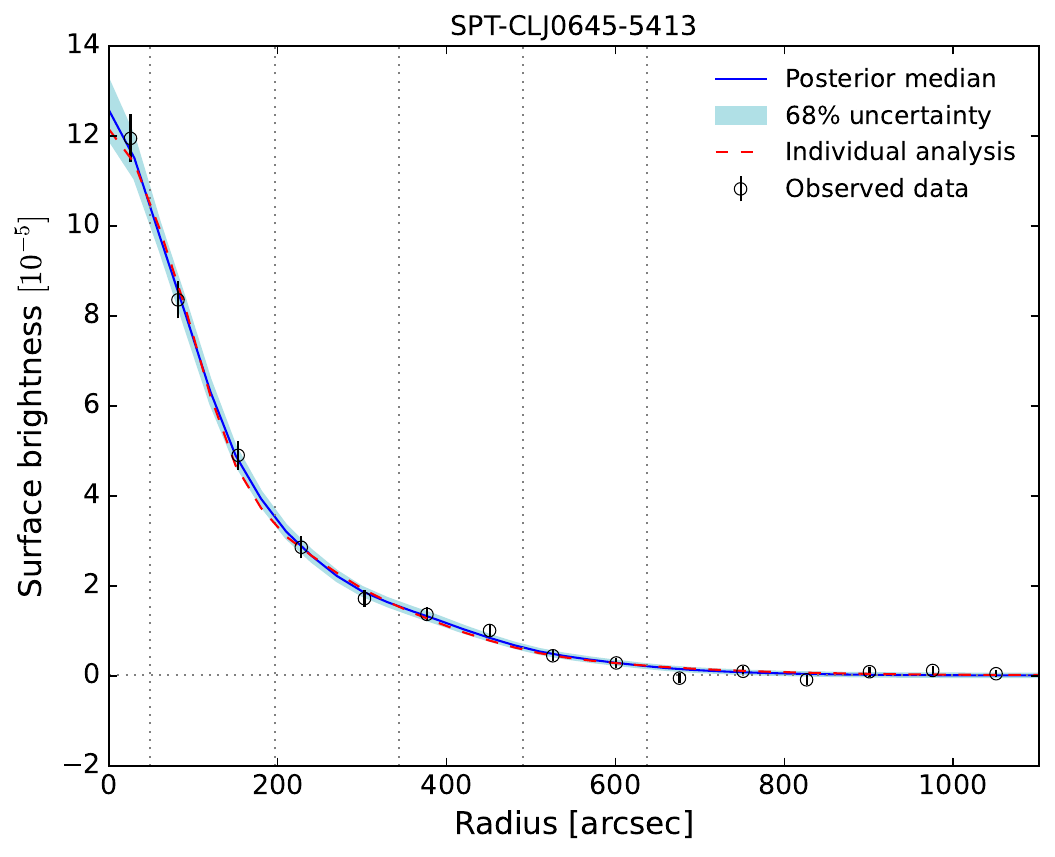}
\includegraphics[width=.48\textwidth, page=2]{plots/fit_on_data.pdf}
\caption{Surface brightness profiles of two clusters, taken from the population analysis. Vertical dotted gray lines show the placement of the knots, whereas the horizontal one indicates the estimated pedestal level. The plots highlight that our radial profile parameterization is adequate for these two clusters. Figure~\ref{fig:chisq} quantifies the fit for the entire sample.}
\label{fig:fit_on_data}
\end{figure}

\begin{figure}[h!]
\begin{center}
\begin{tabular}{c}
\includegraphics[width=.955\linewidth]{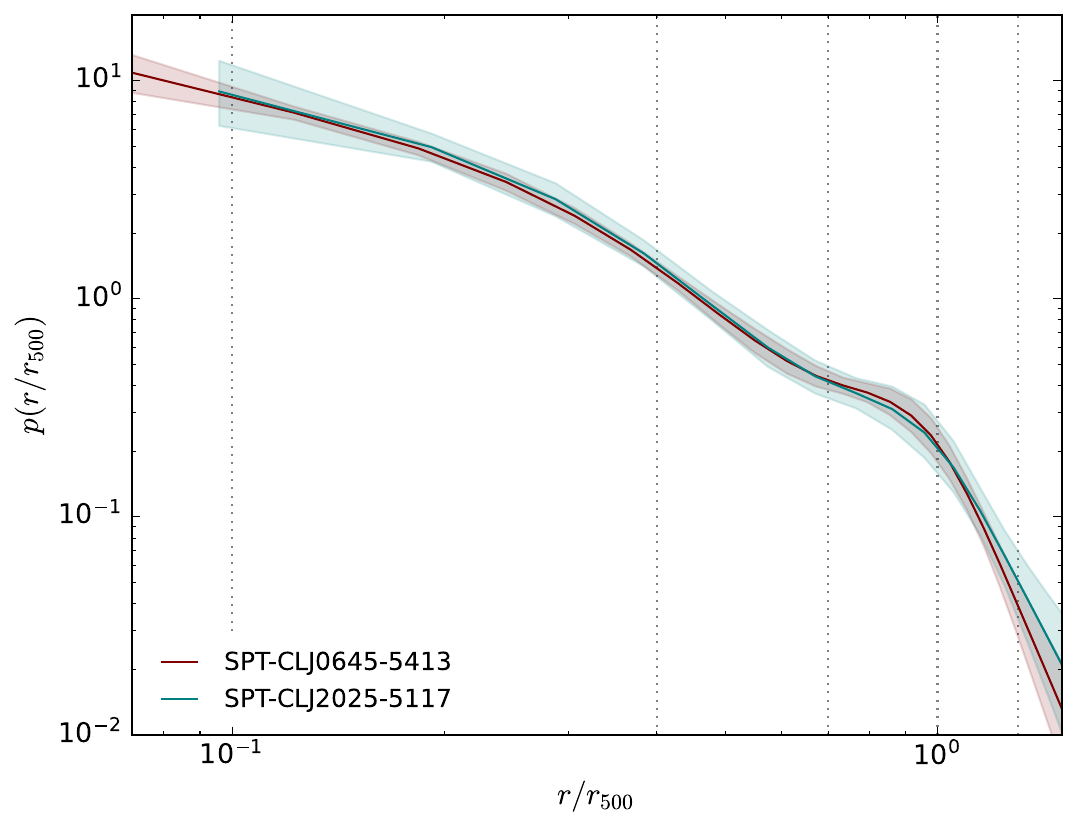}
\end{tabular}
\end{center}
\caption{Pressure profiles of the two clusters in Fig.~\ref{fig:fit_on_data}. Vertical dotted gray lines show the placement of the knots. The plot highlights a marked change in concavity around the third interpolation knot.}
\label{fig:concavity}
\end{figure}

The population analysis simultaneously fits the population-averaged pressure profile and the 55 individual pressure profiles of the galaxy clusters. Figure~\ref{fig:fit_on_data} shows the individual surface brightness profiles from the population analysis for two clusters, taken as examples, together with the observed data and the corresponding profiles from the individual analyses of the previous section.
The plot highlights an acceptable fit of the hierarchical model for the two selected clusters, quantified, as mentioned, in Fig.~\ref{fig:chisq} for all 55 clusters, demonstrating that the chosen model is appropriate for the data used. The two chosen clusters undergo a change in concavity in the pressure profile around the third interpolation knot (see Fig.~\ref{fig:concavity}), an aspect that our model accommodates, unlike other pressure profile parametrizations.

Figure~\ref{fig:literature} compares our individual cluster pressure profile estimates with derivations by other authors who fitted non-parametric models on SZ data~\citep{Ghirardini2019, Sayers2023}, for the four objects for which data are available.
Our pressure profiles are generally consistent with the results from other studies.

Trace plots and marginal probability distributions for the population parameters (color-coded by walker) are shown in Fig.~\ref{fig:traceplot}. 
Table~\ref{tab:population} lists the population parameter estimates (the population-averaged pressures $lgP_k$ and the intrinsic scatters $\sigma_{int,k}$), while Table~\ref{tab:individual} lists the individual pressure parameter estimates $lgP_{k,i}$. These individual estimates are more accurate than those obtained from fitting each cluster independently, as the joint hierarchical model leverages the population-level posteriors as informative priors for the individual $lgP_{k,i}$.

\begin{figure}
\centering  
\begin{center}
\includegraphics[width=.48\textwidth, page=2]{plots/cornerplot.pdf} \\
\includegraphics[width=.48\textwidth, page=1]{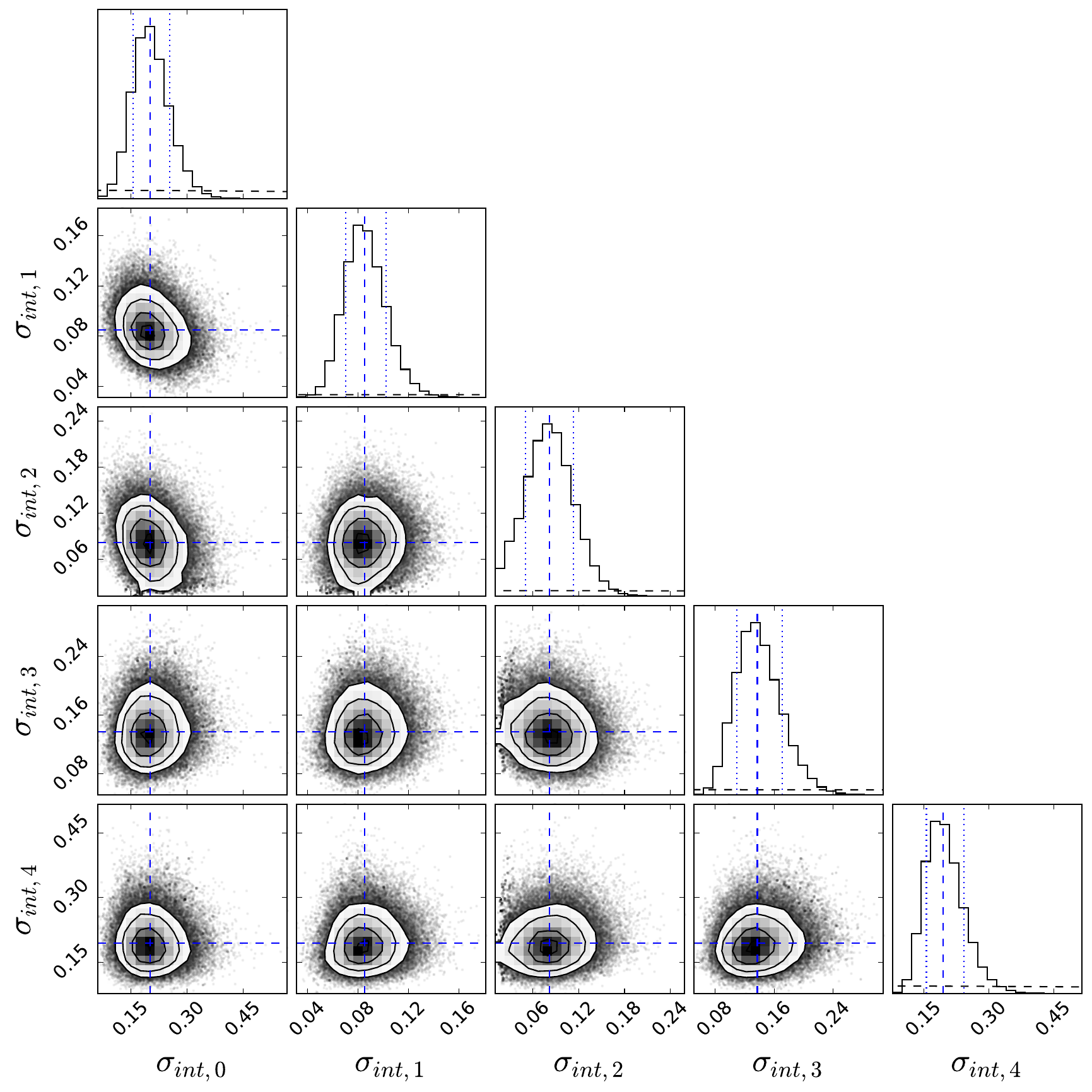}
\caption{Joint and marginal posterior distributions for the population-averaged pressure parameters $lgP_k$ (top panel) and intrinsic scatters $\sigma_{int,k}$ (bottom panel) at the five radii (interpolation knots). Blue dashed lines represent the median values. In the diagonal panels, blue dotted vertical lines indicate the 68\% uncertainty interval, while black dashed lines represent the adopted prior. All parameters show little correlation, due to our choice of the radial profile modelization and the knots spacing. Prior distributions have a negligible impact on the posterior distributions.}
\label{fig:cornerplot}
\end{center}
\end{figure} 

\begin{figure}
\begin{center}
\begin{tabular}{c}
\includegraphics[width=.955\linewidth]{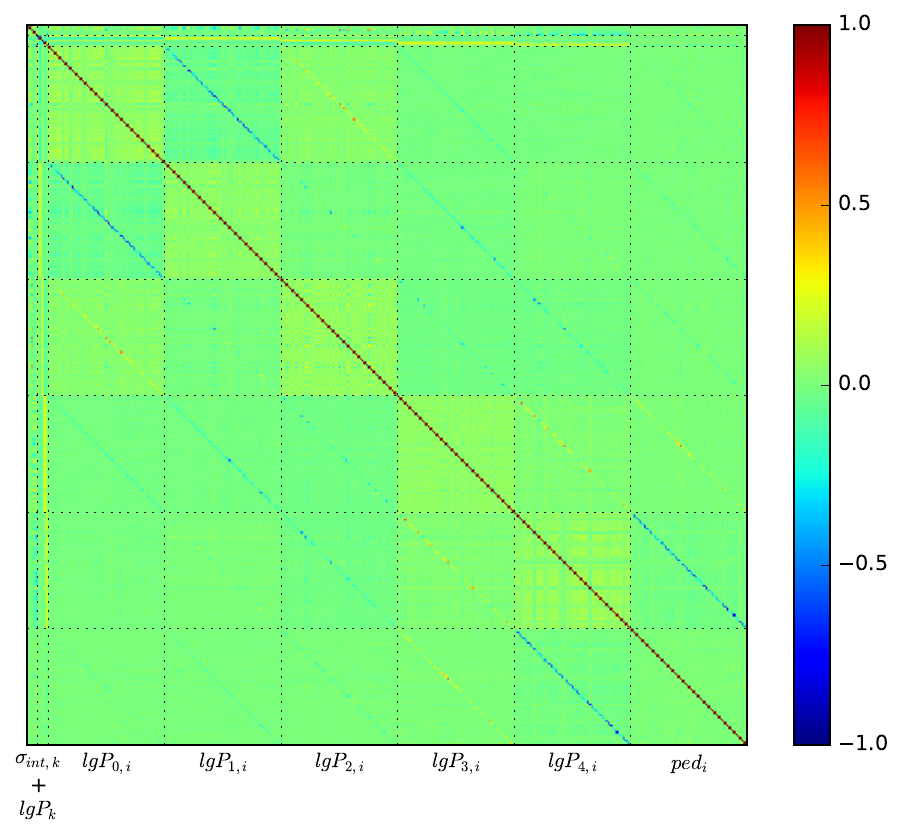}
\end{tabular}
\end{center}
\caption{Correlation matrix for all 340 fitted parameters. Most parameters show little covariance, except between the pressures of the two inner knots and between the pressures at the last knot and the pedestal parameters (see text for details). These low correlations contrast with those observed in the fit using a gNFW model (see Fig.~\ref{fig:corrmat_gnfw}).}
\label{fig:corrmat}
\end{figure}

Figure~\ref{fig:cornerplot} shows the joint and marginal probability distributions for the population parameters: the population-averaged pressure values (top panel) and the intrinsic scatters (bottom panel) at the five radii $\left[ 0.1, 0.4, 0.7, 1, 1.3 \right] \times r_{500}$.
The Bayesian approach inherently accounts for the covariance among parameters. However, this covariance is weak for these parameters due to our choice of the modeling function (a restricted cubic spline) and the spacing of the knots.
Figure~\ref{fig:corrmat} shows the correlation matrix for all 340 fitted parameters, where each value $\rho_{i,j}$ represents the correlation between the parameter pair ($\theta_i, \theta_j$). While most parameters are nearly uncorrelated, two pairs stand out. The pressure at the outermost radius is strongly anticovariant with the pedestal, indicating that the data cannot distinguish a flat cluster profile from a flat background, despite our prior on the outer slope of the profile. Additionally, the pressures at the first two knots are anticorrelated, as the data largely constrain a quantity near their sum because of PSF effects. As shown in Sect.~\ref{sec:conc_stat}, our modelization for the pressure is preferable to using a gNFW profile.
Since the prior distributions are nearly uniform in the range where the posteriors are non-zero (Fig.~\ref{fig:cornerplot}), their impact on estimating the population parameters is minimal. 
Even the prior constraint to force a negative outer slope (beyond $1.3r_{500}$) has a negligible impact on the results: the $2\sigma$ upper limit of the slopes is $<0$ for 51 out of 55 clusters and the population-averaged mean slope is $-5.38\pm0.65$.
We also performed an alternative fit, modeling the inverse of the intrinsic scatter with a Gamma prior distribution, and obtained similar results.

\begin{figure}
\begin{center}
\begin{tabular}{c}
\includegraphics[width=.955\linewidth]{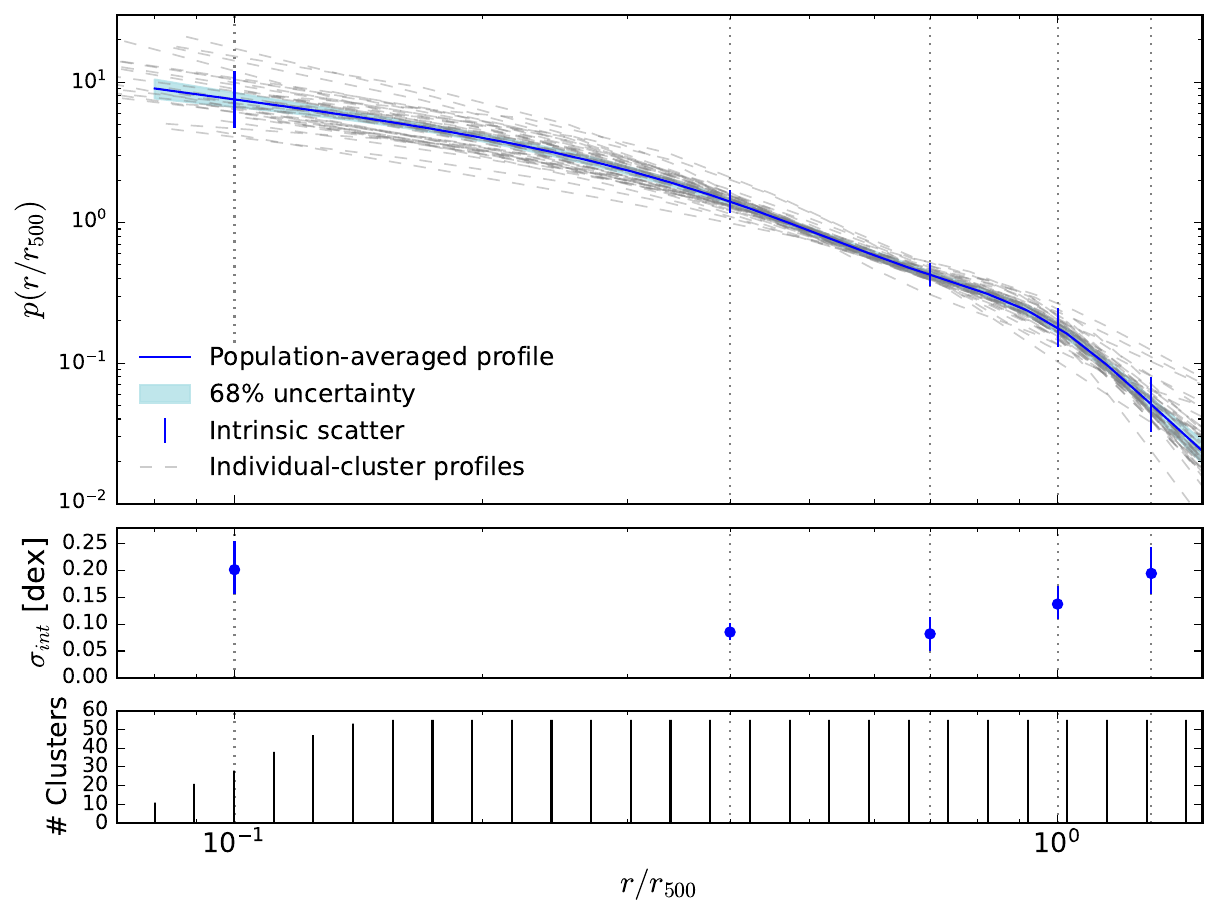}
\end{tabular}
\end{center}
\caption{Pressure profiles (top panel) and intrinsic scatter (top and central panels) of our population analysis. The scatter is minimal at $0.4r_{500}\leq r \leq0.7r_{500}$. The bottom panel shows the number of objects contributing at each radius.}
\label{fig:press_joint}
\end{figure}

Figure~\ref{fig:press_joint} shows the main result of our work encompassing estimates from both levels of the hierarchical model.
The top panel shows the population-averaged pressure profile (solid blue line), its 68\% uncertainty (blue shaded area), the 68\% intrinsic scatter around it (error bars), and the individual cluster estimates from the population analysis (gray dashed lines). 
We emphasize that, with noisy data, we expect more than 68\% of the individual posterior median profiles to lie within the 1-$\sigma_{int}$ interval. This occurs because the prior on the population parameters attracts the individual posterior parameters toward the mean, as in the SN fit shown by \citet{Andreon2013}.
The central panel shows the radial distribution of the intrinsic scatter, estimated at the five knots. The dispersion is minimal at intermediate radii (second and third knots), which confirms and reinforces that the intermediate cluster regions are the most regular~\citep{Arnaud2010}.
At an attentive inspection, at least some individual profiles (e.g., SPT-CLJ0645-5413, SPT-CLJ2025-5117, SPT-CLJ0225-4327) undergo a change in concavity between the third and fourth knot (as shown in Fig.~\ref{fig:concavity}). The fraction of profiles with a change in concavity depends on the threshold used to classify objects as members of the class.
This change in concavity is not allowed with a gNFW profile, and as a consequence, has never been seen in previous works using this radial model. The inspection of the data (e.g., Fig.~\ref{fig:fit_on_data}) confirms that the model captures the data behavior. Inspection of the XMM X-ray data for the clusters exhibiting a change in concavity does not reveal any obvious distinguishing features compared to the rest of the sample.
The bottom panel shows the number of clusters that contributed to the derivation of the population-averaged pressure profile at different radii, noting that for some cluster the innermost radial bin (30 arcsec) is larger than the innermost interpolation knot. Around half of the clusters in the sample contribute to the determination of the intrinsic scatter at the innermost interpolation knot, $r=0.1r_{500}$.

\begin{figure}
\begin{center}
\begin{tabular}{c}
\includegraphics[width=.955\linewidth]{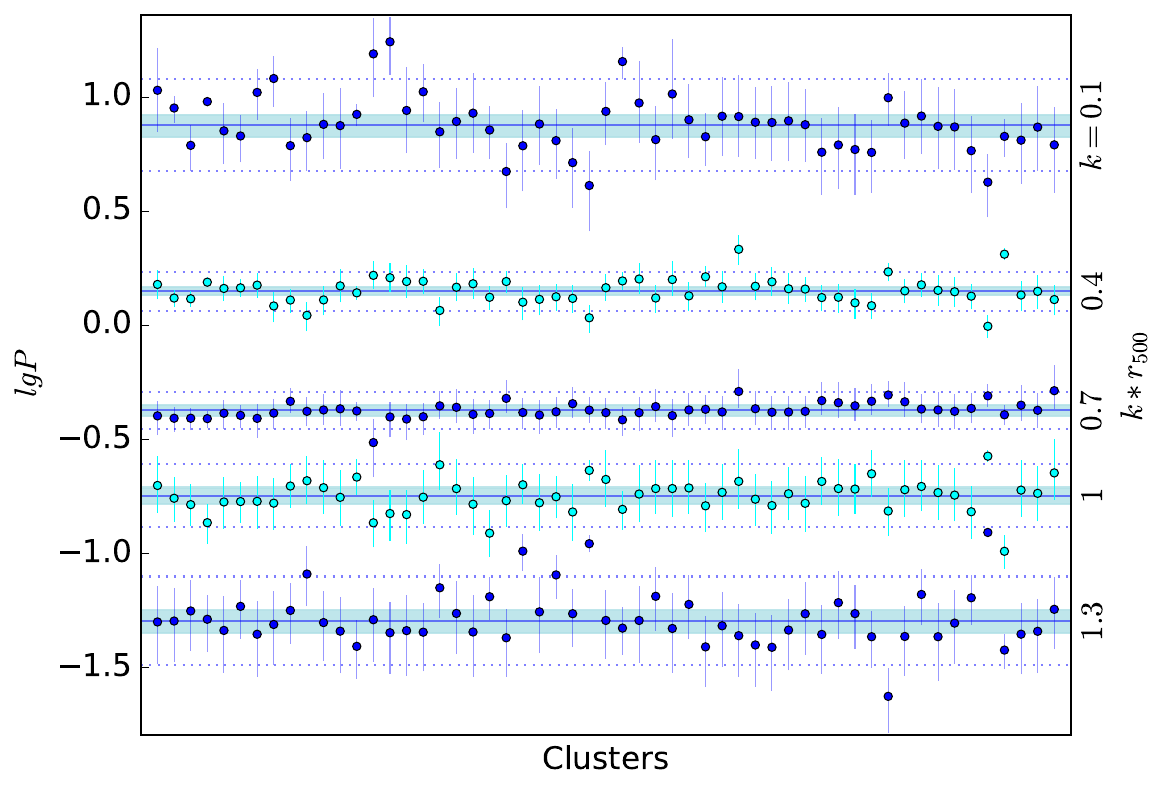}
\end{tabular}
\caption{Individual pressure parameters $lgP_{k,i}$ (points with error bars) at the five knots (marked on the right y-axis). Horizontal lines represent the population-averaged pressure estimates $lgP_k$ (solid lines) with their errors (shaded areas), and intrinsic scatter estimates (dotted lines).}
\label{fig:outliers}
\end{center}
\end{figure}

\begin{figure}
\begin{center}
\begin{tabular}{c}
\includegraphics[width=.955\linewidth]{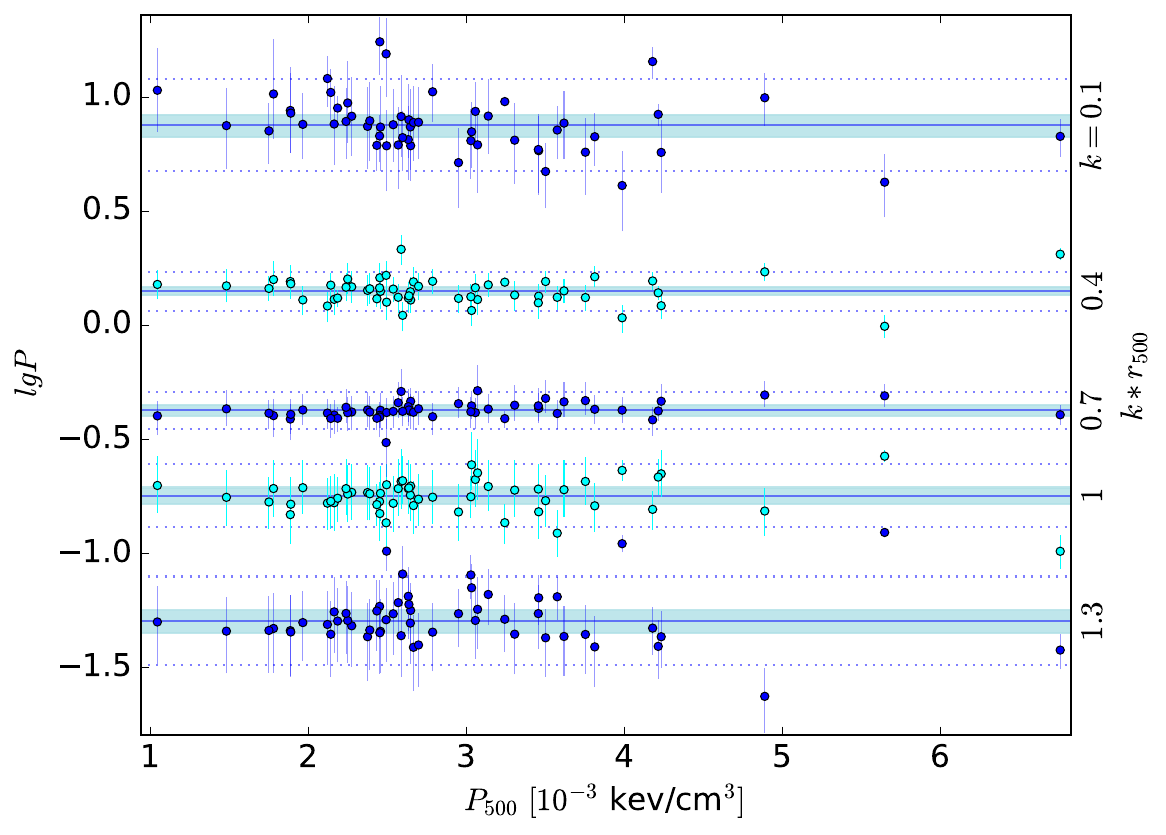}
\end{tabular}
\caption{Individual pressure parameters $lgP_{k,i}$ (points with error bars) at the five knots (marked on the right y-axis) as a function of the cluster's $P_{500}$. Horizontal lines represent the population-averaged pressure estimates $lgP_k$ (solid lines) with their errors (shaded areas), and intrinsic scatter estimates (dotted lines).}
\label{fig:P500_analysis}
\end{center}
\end{figure} 

Figure~\ref{fig:outliers} shows the pressure estimates of individual clusters $lgP_{k,i}$ (points with error bars) together with the population-averaged pressure values $lgP_k$ (solid lines), their errors (shaded areas) and intrinsic scatters (dotted lines).
This visualization highlights that we did not detect any cluster with outlier pressure parameters, despite allowing for such outliers by using a Student-t distribution to model the scatter. Only two individual pressure estimates lie beyond the 2-$\sigma_{int}$ threshold: one at the second and one at the fifth knot.
Figure~\ref{fig:P500_analysis} shows the pressure estimates of individual clusters as a function of $P_{500}$, suggesting that there is no discernible trend of dependence on it.

\begin{figure}
\begin{center}
\begin{tabular}{c}
\includegraphics[width=.9\linewidth]{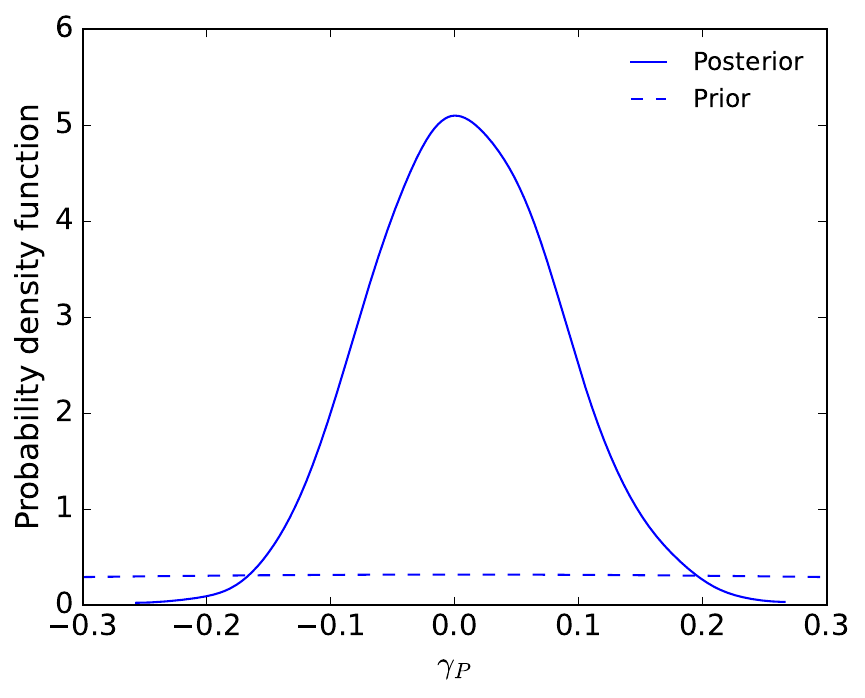}
\end{tabular}
\end{center}
\caption{Probability density function for the posterior and prior distributions of $\gamma_P$. A Bayes factor of 16.5 is moderately in favor of no dependence on $P_{500}$.}
\label{fig:bf}
\end{figure}

To formally evaluate potential dependencies of the population parameters 
on $P_{500}$ and redshift, we extended our baseline fitting model to include these effects. 
Specifically, we added to Eq.~\ref{eq:lgpki}, in turn, $\gamma_P \log (P_{500,i}/P_{500,med})$ and $\gamma_z \log ((1+z_i)/(1+z_{med}))$, where $P_{500,med}$ and $z_{med}$ are the sample medians.
We adopted a Student's $t$ distribution with one degree of freedom (equivalent to a uniform prior on the angle) for each additional parameter $\gamma$. To assess the preference for the extended model over the baseline one, we computed the Bayes factor using the Savage-Dickey density ratio~\citep{dickey1971}, as the models are nested.
As shown in Fig.~\ref{fig:bf}, the data favor the absence of a dependency on $P_{500}$, with a Bayes factor of 16.5, indicating moderate to strong evidence and supporting our decision not to include this term in the model. Instead, the comparison yields inconclusive evidence for redshift dependence, with the relative odds between the models being comparable.

\begin{figure}
\begin{center}
\begin{tabular}{c}
\includegraphics[width=.955\linewidth]{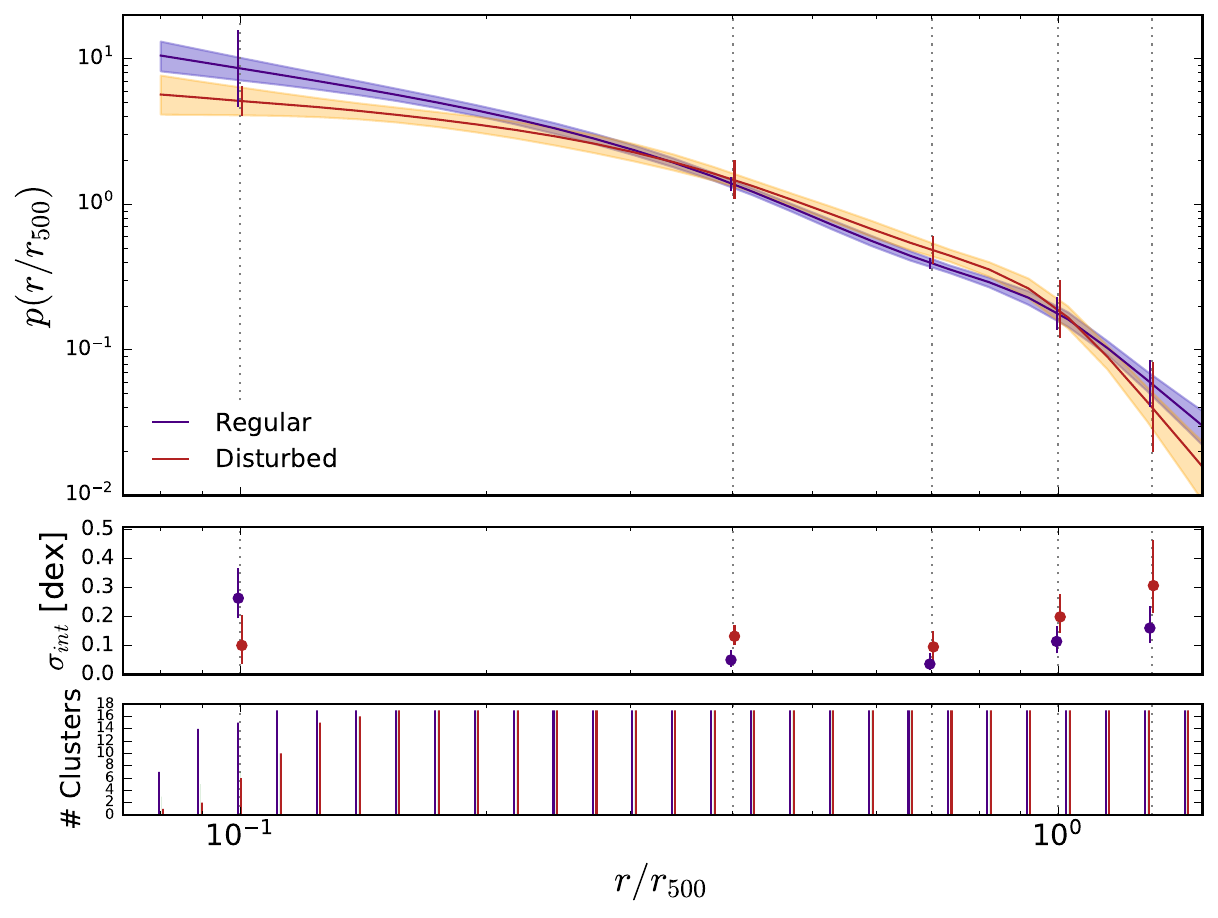}
\end{tabular}
\caption{Pressure profiles (top panel) and intrinsic scatter (top and central panels) of regular and disturbed clusters. Error bars and shading as in Fig.~\ref{fig:press_joint}. Points are slightly horizontally offset for clarity. Regular clusters seem to show a lower scatter at $r\geq0.4r_{500}$ and a more peaked profile at the center. 
The bottom panel shows the number of objects contributing at each radius.}
\label{fig:morpho}
\end{center}
\end{figure}

Figure~\ref{fig:morpho} shows the population-averaged pressure profiles of the regular and disturbed clusters separately. These profiles were derived from joint fits of clusters belonging to these classes, as done for the entire sample, and the estimated parameters are reported in Table~\ref{tab:population}. Regular clusters exhibit a more peaked profile at the center (indicative of cool cores) compared to disturbed clusters, and show lower intrinsic scatter beyond the innermost knot, indicating that part of the pressure profile heterogeneity in the full sample is related to cluster morphology.
The pressure profile and scatters of clusters morphologically classified as intermediate lie between them (not shown in the plot to avoid crowding). Since the disturbed clusters subset includes objects with a disturbance indicator spanning a large range ($0.4<D_{\text{comb}}\le1$), we also performed an additional comparison by splitting this group into two further subsets, using $D_{\text{comb}}=0.75$ as the threshold. Even in this case, no significant differences were found.

\begin{figure}
\begin{center}
\begin{tabular}{c}
\includegraphics[width=.955\linewidth]{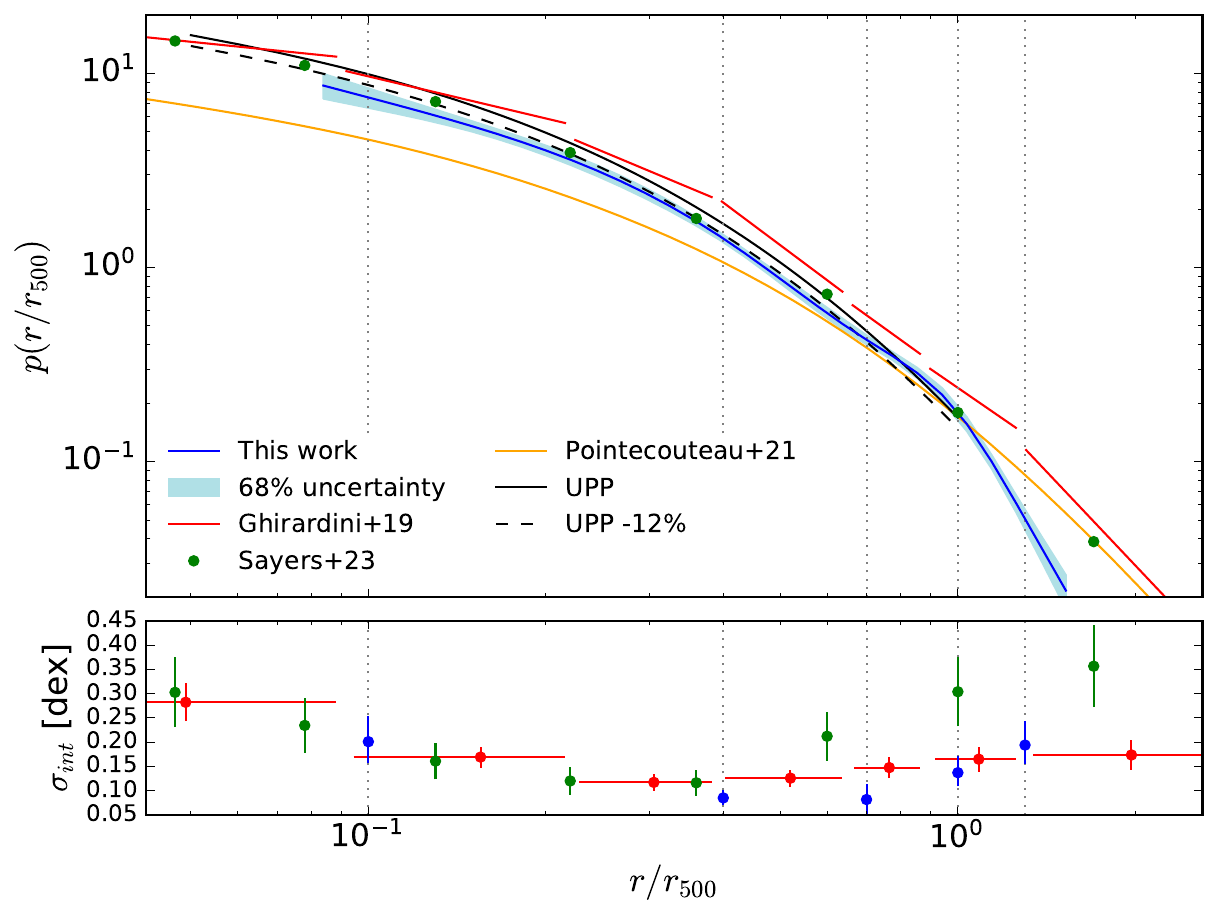}
\end{tabular}
\caption{Pressure profiles (top panel) and intrinsic scatters (bottom panel) for our sample and literature works, with literature errors omitted from the top panel for clarity. Within $r_{500}$, our profile is $\sim$12\% lower than the UPP, close to what is needed to remove the tension in cosmological parameters derived from CMB and \textit{Planck} SZ cluster counts~\citep{Ruppin2019}. The steep decline of our pressure profile beyond $r_{500}$ is consistent with the deficit reported by \citet{Anbajagane2022}, interpreted as a shock-induced thermal non-equilibrium between electrons and ions. Regarding the scatter, the three radial profiles are consistent up to $0.4r_{500}$. Beyond this radius, \citet{Sayers2023} report the highest scatter, while our analysis yields the lowest.}
\label{fig:lit_comp}
\end{center}
\end{figure}

The top panel of Fig.~\ref{fig:lit_comp} compares our population-averaged pressure profile with results from other works. 
Our profile exhibits a pressure deficit of approximately $12\pm3\%$ compared to the UPP~\citep[][black solid line]{Arnaud2010}. The median absolute fractional difference between our profile and the $12\pm3\%$-reduced UPP (black dashed line) is only $\sim$6\%, indicating that deviations are minimal at the resolution offered by our data (250 kpc FWHM).
Within $r_{500}$, our profile agrees well with that of the low-$z$ sample from \citet[][green points]{Sayers2023}. 
The profile of \citet[][orange line]{Pointecouteau2021} shows an even more marked pressure deficit than ours within $0.6r_{500}$, while that of \citet[][red lines]{Ghirardini2019} lies systematically above all others.
Beyond $r_{500}$, our profile declines more steeply than those in these three works, which mostly rely on \textit{Planck} at large radii. 

The observed $\sim$12\% pressure deficit relative to the UPP is consistent with the $\sim$15\% reduction proposed by \citet{Ruppin2019} to alleviate the tension between cosmological parameters derived from the CMB and \textit{Planck} SZ cluster counts. It is also in agreement with the revised UPP from \citet{He2021}, which assumes a mass bias derived from simulations and whose predicted power spectrum at multipoles $100<l<1000$ approximately matches the value observed by \textit{Planck}, whereas the original UPP overestimated the signal by up to 50\%.
The steep decline of our pressure profile beyond $r_{500}$ is consistent with the deficit reported by \citet{Anbajagane2022}, interpreted as a shock-induced thermal non-equilibrium between electrons and ions. 
This lower pressure and steeper profile beyond $r_{500}$ is also in line with the conditions required to reconcile the observed and predicted angular power spectrum of the thermal SZ effect in the work of \citet{Ramos-Ceja2015}.

As previously noted, our pressure values at the outermost knot are correlated with the pedestal estimates. This correlation does not significantly affect the slope at large radii: removing the pedestal from the model and refitting the data yields a similar population-averaged profile. Nevertheless, distinguishing a weak, approximately constant signal from the background remains inherently challenging, both in our work and in other works, regardless of whether the pedestal is included in the model.

Sample selection is unlikely to explain the differences among the pressure profiles obtained by different works. \citet{Pointecouteau2021} and \citet{Ghirardini2019} analyzed SZ-selected samples, like ours, yet found discrepant results in opposite directions, with the \citet{Ghirardini2019} profile lying above the UPP, potentially increasing the cosmological tension. \citet{Arnaud2010} and \citet{Sayers2023} studied X-ray-selected samples - based on XMM and Chandra data, respectively - with \citet{Sayers2023} also incorporating high-resolution SZ observations. Pressure estimates including X-ray data are less direct due to the assumption of clumpiness~\citep{Eckert2015} and differences between mass-weighted and X-ray temperatures~\citep{Vikhlinin2006}. 
Additionally, a long-standing issue with absolute X-ray calibration, particularly between Chandra and XMM~\citep{Schellenberger2015}, may contribute to the observed differences when not considered.
Since all of these works probe comparable regions of the redshift-$P_{500}$ plane, and we detect no dependence of the pressure parameters on redshift or $P_{500}$ (Fig.~\ref{fig:bf}), the observed differences between profiles cannot be attributed to variations in either of these parameters.

The bottom panel of Fig.~\ref{fig:lit_comp} compares our estimate of the intrinsic scatter with those of \citet{Ghirardini2019} and \citet{Sayers2023}, both of which, like our work, adopt a non-parametric model for the pressure profile. 
The three radial profiles of intrinsic scatter are consistent up to $0.4r_{500}$. Beyond this radius, \citet{Sayers2023} report the highest scatter, while our analysis yields the lowest.
Although SZ-selected samples are expected to encompass more heterogeneous populations of clusters than X-ray-selected samples~\citep{Planck2011IX, Andreon2016, Rossetti2017}, this expectation is not reflected in the intrinsic scatter estimates: the scatter in the X-ray-selected sample of \citet{Sayers2023} is systematically larger beyond $0.6r_{500}$ than in the SZ-selected samples of our work and \citet{Ghirardini2019}. 

\subsection{Sensitivity analyses}

Our analysis excludes three clusters within the redshift range, potentially introducing some incompleteness into the sample. Although formally incomplete, the sample remains representative because these clusters were removed for reasons not related to their Compton-$y$ signal, making it a random sampling of a complete cluster, and therefore their removal should not affect our results. Nevertheless, we assess the impact of this random incompleteness by slightly narrowing the redshift range and re-fitting all 49 clusters within the interval $0.08 < z < 0.29$, a range in which no clusters were excluded. The resulting posterior means and uncertainties of the population-averaged pressure parameters at the five knots, along with three of the five intrinsic scatter parameters, remain unchanged. The posterior means of the remaining two intrinsic scatters differ by approximately 1-$\sigma$.

Our analysis includes clusters with Compton-$y$ S/N as low as 5, consistent with other studies that also adopt similar or even lower thresholds~\citep[e.g.,][]{Pointecouteau2021}. To assess the robustness of our results to the S/N cut, we repeated the analysis using only the 33 clusters within our sample with $S/N>6$. The results are very similar to those obtained from the full sample, with a mean absolute standardized difference of 0.56 between the population parameters of the two models.

In our analysis, we adopted a specific Y-M scaling relation (Eq.~\ref{eq:Y500-M500}). Preliminary tests using an alternative scaling relation and a reduced sample, chosen to limit computational cost, indicate that the resulting mean pressure profile is sensitive to the choice of scaling relation. The observed variation exceeds our current uncertainty estimates, due to the precision achieved by our large sample. As with all analyses, our results are conditional on the underlying assumptions, which in our analysis and similar ones include the adopted cosmology and the Y-M scaling relation. A more comprehensive investigation, possibly including a simultaneous fit of the scaling parameters, is deferred to future work.

\section{Discussion and conclusions} \label{sec:discussion}

\subsection{Astrostatistics considerations} \label{sec:conc_stat}

\begin{figure}
\begin{center}
\begin{tabular}{c}
\includegraphics[width=.955\linewidth]{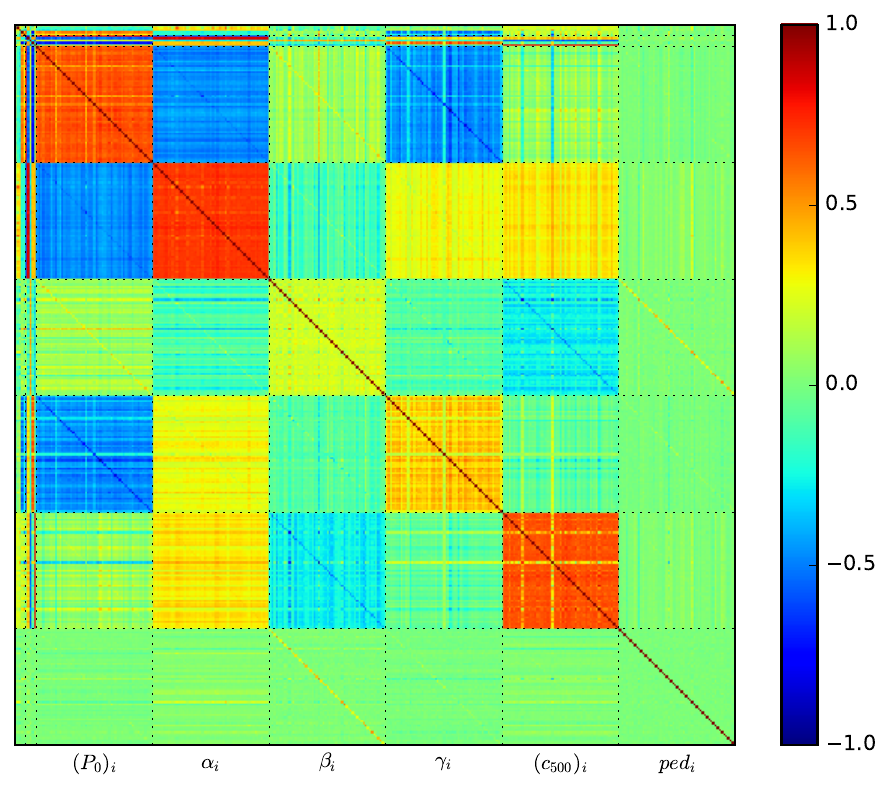}
\end{tabular}
\end{center}
\caption{Correlation matrix for all 340 fitted parameters in an analysis performed assuming a gNFW model for the pressure profile. Many parameters show a strong level of covariance, much higher than with our pressure parameterization, whose correlation matrix is shown in Fig.~\ref{fig:corrmat}.}
\label{fig:corrmat_gnfw}
\end{figure}

Our approach sets a new standard compared to previous analyses. 
First, we introduce the restricted cubic spline interpolation model for the pressure profile, which offers several advantages. It improves on the cubic spline interpolation model~\citep{Andreon2021, Andreon2023} as it avoids undesired twists in the pressure profile at radii that are essentially unconstrained by the data. 
It does not show the irregular uncertainties (larger near the knots and smaller at the center of the interpolation intervals) observed with the power-law interpolation model adopted, for example, by \citet{Romero2018} and \citet{keruzore2023}. By not assuming a parametric form for individual cluster and population-averaged profiles, we do not impose a model on nature, unlike the gNFW radial profile. As a result, the restricted cubic spline model can flexibly capture changes in the concavity of the pressure profile, unlike more constraining parametrizations such as the gNFW or power-law models.
Second, our choice of profile parameterization makes the parameters easier to interpret and breaks the strong covariance present, for example, in fits that assume a gNFW model. Figure~\ref{fig:corrmat_gnfw} shows the correlation matrix for the analysis performed using a gNFW parameterization~\citep{Nagai2007} instead of restricted cubic splines. The differences compared to the correlation matrix in Fig.~\ref{fig:corrmat} are substantial.
Third, by fitting the population and the individual clusters simultaneously in a BHM, we avoid the inconsistency of widespread two-step analyses, which first assume a uniform prior for a parameter (e.g., the profile normalization) during the individual fits and subsequently impose a normal distribution on it.
Fourth, by adopting a Student-t distribution, we account for outliers, i.e., clusters with unusual pressure profiles, which are automatically identified and downweighted in the computation of the average profile. This approach eliminates the need for manual examination and the use of arbitrary criteria (such as a $3\sigma$ clipping, which may be based on a preliminary computed $\sigma$), as previously done~\citep[e.g.,][]{Ghirardini2019}. 
Fifth, rather than blindly assuming a model for the individual pressure profile, we critically assess its validity by checking how well it fits the data (see Fig.~\ref{fig:chisq} and discussion therein). Similarly, rather than blindly assuming no evolution and no $P_{500}$ dependency of population parameters, we use the Bayes factor to choose whether to include them in the model.
Finally, when a parameter is unconstrained, rather than arbitrarily fixing its value and excluding all other possibilities, including those differing negligibly, our joint fit marginalizes over the prior distribution. When a parameter is unconstrained for a cluster, the joint fit uses the information from the whole sample as an effective prior, improving its value and uncertainty.

From a computational perspective, this process is highly time-consuming due to the simultaneous fitting of 340 parameters.
However, PyMC is particularly effective for modeling complex dependencies in hierarchical Bayesian frameworks. It supports multidimensional parameters in hierarchical models by allowing group-level distributions to define structured dependencies across multiple levels. PyMC also enables vectorized priors through the \textit{shape} argument, ensuring computational efficiency and automatic sharing of statistical strength across groups.

\subsection{Astronomical considerations} \label{sec:conc_astro}

We fitted our model on the largest non-stacked sample of individual cluster radial profiles (55 objects), computing the population-averaged profile and the intrinsic scatter around it, as described by the parameters in Table~\ref{tab:population}. We also derived the parameters of the individual pressure profiles (see Table~\ref{tab:individual}) by leveraging information from clusters with well-determined profiles.
By applying appropriate selection criteria in mass and redshift, we avoid attempting to measure the pressure profile of clusters that are unresolved with the data (see Fig.~\ref{fig:catalogs}).
Following the identification of inconsistencies in some $M_{500}$ estimates, we rederived all normalizations under weaker assumptions. We identified some clusters with radial pressure profiles that exhibit a change in concavity. Although the model allows for outliers, none are detected in the pressure values of individual clusters.
Compared to \citet{Arnaud2010}, our determination extends to 1.4x larger radii and includes more objects in each morphologically-defined class (17 vs 10 or 11). Compared to \citet{Ghirardini2019}, whose scatter estimates are based on just 4 and 8 clusters per class and thus considered potentially unreliable, we can split our sample into morphological classes with a significantly larger number of objects per class. 
Despite the unprecedented precision enabled by the largest sample of resolved clusters and the flexibility offered by our spline-based radial profile, the population-averaged profile is close to the UPP within $r_{500}$ (Fig.~\ref{fig:press_joint}), indicating that deviations are minimal at the resolution offered by our data (250 kpc FWHM). Our profile is even more closely aligned with the $\sim$15\% decreased version of the UPP proposed by \citet{Ruppin2019} to alleviate tensions between cosmological parameters as derived from the CMB and \textit{Planck} SZ cluster counts.
The steep decline of our pressure profile beyond $r_{500}$ is consistent with the deficit reported by \citet{Anbajagane2022}, interpreted as a shock-induced thermal non-equilibrium between electrons and ions. 
At large radii, specifically around $r = 1.4r_{500}$, distinguishing a weak, approximately constant signal from the background is inherently challenging. As a result, pressure measurements and slope estimates at these radii should be interpreted with caution. This is true regardless of whether the background level along the cluster line of sight is allowed to deviate from that measured in a reference region (as done in our analysis) or is assumed to be identically zero (as in several other studies).
The dependencies on redshift and $P_{500}$ are also weak and not appreciable in our sample.
The intrinsic scatter is minimal at intermediate radii, approximately between $0.4r_{500}$ and $0.7r_{500}$, and is lower than that determined by \citet{Sayers2023}, who use X-ray-selected samples.
This result is especially notable given that our sample is SZ-selected, and thus expected to encompass a more heterogeneous population of clusters.
Splitting the sample according to morphological types gives similar pressure profiles for the two classes, with some evidence that morphologically regular clusters show a lower intrinsic scatter and a more peaked profile at the center than disturbed clusters.
Cluster morphology is a component contributing to the heterogeneity of pressure profiles within the analyzed sample.

\subsection{Limitations and future developments}

Samples without any SZ-based selection do not require corrections for a never-applied SZ selection. Therefore, our approach can be directly applied - without modification - to samples selected, for example, through gravitational lensing, such as those detected by \citet{Oguri2021} and studied by \citet{Andreon2025}, or those to be released in the upcoming Data Release 1 of the Euclid survey~\citep{Mellier2025}.
However, for samples like the one analyzed here, as well as in most previous works, the selection function has to be modeled and incorporated into the fit. Therefore, the conclusions derived from our analysis apply only to the studied sample or to samples selected in similar ways, rather than to the entire population of clusters. As a consequence, the heterogeneity observed in our sample is an underestimate of the actual heterogeneity. Indeed, there are already known examples of clusters whose pressure profile is manifestly different from those derived from ICM-selected samples~\citep{Andreon2019, Andreon2021, Andreon2023}. Neglecting the selection function was a necessary simplification in this first step; modeling it represents a natural direction for future work. 
Additional possible developments include the use of multi-wavelength data, such as combining SZ measurements with X-ray and weak lensing observations, which would further enhance the robustness of our analysis by integrating complementary information and providing a more complete picture of cluster properties.

\section{Conclusions} \label{sec:conclusion}

We developed a Bayesian hierarchical model that simultaneously and coherently fits both the parameters of individual clusters and those of the population from which they are drawn.
We introduced a highly flexible, low-covariance parameterization of the pressure profile using restricted cubic splines, which provide a simple and intuitive interpretation of the parameters. To account for outliers, we modeled the scatter with a Student-t distribution and assessed the appropriateness of the fitted radial profile against the data. We also applied the beam and transfer function corrections, as required for SZ data. We use the Bayes factor to choose between possible modelizations.
Our method was applied to the largest non-stacked sample of clusters with resolved pressure profiles, extracted from SPT+\textit{Planck} Compton-$y$ maps. This sample consists of 55 clusters, large enough to be divided into three morphological classes based on eROSITA data. Computations are feasible within a few days on a modern (2024) personal computer.
The shape of the population-averaged pressure profile, at our 250 kpc resolution, closely resembles the universal pressure profile, despite the flexibility of our model to accommodate alternative shapes, with a $\sim$12\% lower normalization, close to what is needed to alleviate the tension between cosmological parameters derived from the CMB and \textit{Planck} SZ cluster counts~\citep{Ruppin2019}.
At the largest radii, our profile is steeper than other works, but in agreement with the pressure deficit observed by \citet{Anbajagane2022}, interpreted as a shock-induced thermal non-equilibrium between electrons and ions. A lower pressure profile that steepens at large radii is also required to reconcile the predicted and observed thermal SZ power spectrum.
The estimate of the intrinsic scatter is consistent with or lower than previous estimates, although our SZ-selected sample is expected to include more heterogeneous clusters than the X-ray-selected samples used for comparison. 
Using our flexible pressure model, we identified a few clusters whose radial pressure profiles display a change in concavity. Although the model allows for outliers, none are detected in the pressure values of individual clusters.
When splitting the sample by morphological type, we found that the profiles of the two extreme tertiles are remarkably similar, with some evidence suggesting that morphologically regular clusters exhibit lower intrinsic scatter and more centrally peaked profiles compared to disturbed clusters. Therefore, morphology is a key contributor to the observed heterogeneity in pressure profiles.

\section*{Data availability}
Full Table~\ref{tab:individual} is only available in electronic form at the CDS via \url{http://cdsweb.u-strasbg.fr/cgi-bin/qcat?J/A+A/}.

\begin{acknowledgements}
We thank Merrilee Hurn for helpful statistical advice, Om Sharan Salafia for fruitful discussion, and Jack Sayers for useful comments.
FC and SA acknowledge INAF grant "Characterizing the newly discovered clusters of low surface brightness" and PRIN-MIUR grant 20228B938N "Mass and selection biases of galaxy clusters: a multi-probe approach", the latter funded by the European Union Next generation EU, Mission 4 Component 1 CUP C53D2300092 0006.
\end{acknowledgements}

\bibliographystyle{aa} 
\bibliography{biblio}

\begin{appendix}
\onecolumn

\section{Comparison with individual profiles}

\begin{figure*}[h!]
\begin{center}
\begin{tabular}{cc}
\includegraphics[width=.4\textwidth, page=1]{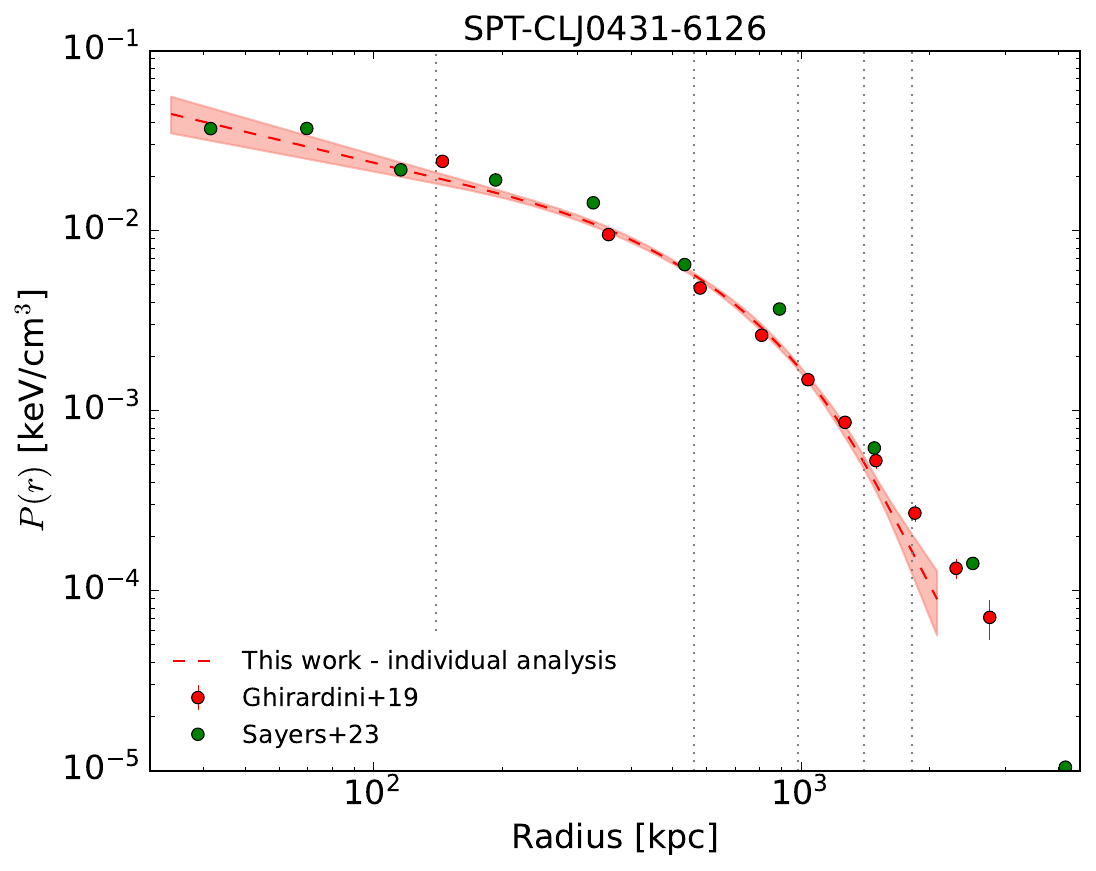} & 
\includegraphics[width=.4\textwidth, page=2]{plots/press_compare_nognfw.pdf}
\\
\includegraphics[width=.4\textwidth, page=3]{plots/press_compare_nognfw.pdf} & 
\includegraphics[width=.4\textwidth, page=4]{plots/press_compare_nognfw.pdf}
\end{tabular}
\caption{Pressure profiles in our work and literature works providing non-parametric determinations using SZ data~\citep{Ghirardini2019, Sayers2023}. Errors on the estimates from other works are shown when available. Vertical gray dotted lines show the placement of the knots. Our profiles show a generally good agreement with results from other authors using independent data.}
\label{fig:literature}
\end{center}
\end{figure*}

\section{Population analysis diagnostics plots}

\begin{figure*}[h!]
\begin{center}
\begin{tabular}{cc}
\includegraphics[width=.47\linewidth, page=1]{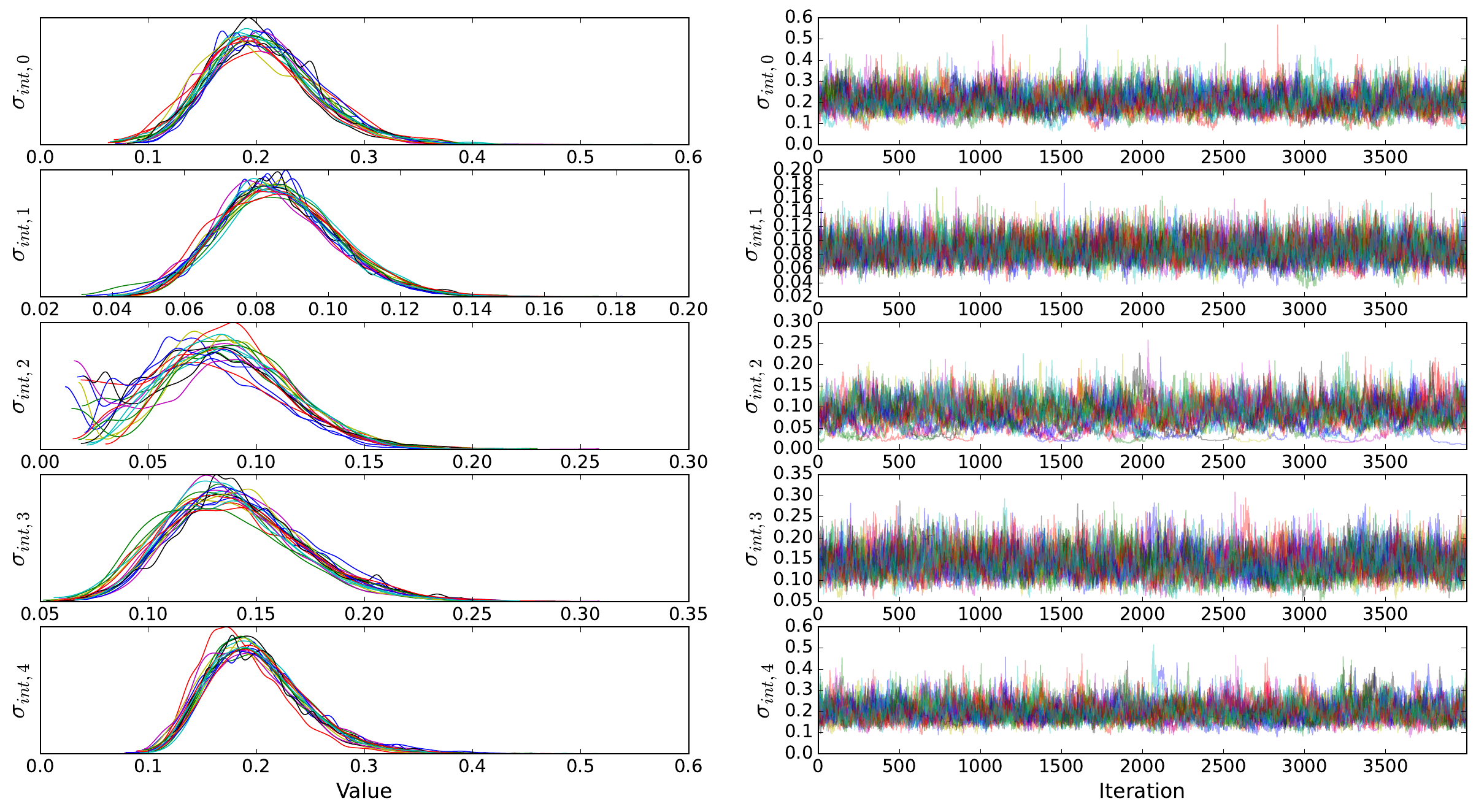} &
\includegraphics[width=.47\linewidth, page=2]{plots/traceplot.pdf}
\end{tabular}
\end{center}
\caption{Marginal posterior distribution (kernel density estimate) and trace plots of the population-level parameters. Each line refers to a different walker in the MCMC.}
\label{fig:traceplot}
\end{figure*}

\end{appendix}
\end{document}